\documentclass[twocolumn,twocolappendix]{aastex63}
\hypersetup{linkcolor=cyan,citecolor=cyan,filecolor=green,urlcolor=blue}

\usepackage{listings}
\usepackage{soul}

\received{November 15, 2019}
\revised{February 5, 2020}
\accepted{April 12, 2020}
\submitjournal{ApJ}

\shorttitle{Star-gas misalignment in galaxies: I}
\shortauthors{D. J. Khim et al.}
\graphicspath{{./}}

\begin{document}

\title{Star-gas misalignment in galaxies: I. the properties of galaxies from the Horizon-AGN simulation and comparisons to SAMI}

\correspondingauthor{Sukyoung K. Yi}
\email{E-mail: yi@yonsei.ac.kr}

\author{Donghyeon J. Khim}
\affiliation{Department of Astronomy, Yonsei University, Seoul 03722, Republic of Korea}

\author{Sukyoung K. Yi}
\affiliation{Department of Astronomy, Yonsei University, Seoul 03722, Republic of Korea}

\author{Yohan Dubois}
\affiliation{Institut d’ Astrophysique de Paris, Sorbonne Universités, et CNRS, UMP 7095, 98 bis bd Arago, 75014 Paris, France}

\author{Julia J. Bryant}
\affiliation{Sydney Institute for Astronomy (SIfA), School of Physics, The University of Sydney, NSW 2006, Australia}
\affiliation{Australian Astronomical Optics, AAO-USydney, School of Physics, University of Sydney, NSW 2006, Australia}
\affiliation{ARC Centre of Excellence for All Sky Astrophysics in 3 Dimensions (ASTRO 3D)}

\author{Christophe Pichon}
\affiliation{Institut d’ Astrophysique de Paris, Sorbonne Universités, et CNRS, UMP 7095, 98 bis bd Arago, 75014 Paris, France}
\affiliation{Korea Institute of Advanced Studies (KIAS), 85 Hoegiro, Dongdaemun-gu, Seoul, 02455, Republic of Korea}

\author{Scott M. Croom}
\affiliation{Sydney Institute for Astronomy (SIfA), School of Physics, The University of Sydney, NSW 2006, Australia}
\affiliation{ARC Centre of Excellence for All Sky Astrophysics in 3 Dimensions (ASTRO 3D)}

\author{Joss Bland-Hawthorn}
\affiliation{Sydney Institute for Astronomy (SIfA), School of Physics, The University of Sydney, NSW 2006, Australia}

\author{Sarah Brough}
\affiliation{School of Physics, University of New South Wales, NSW 2052, Australia}

\author{Hoseung Choi}
\affiliation{Institute of Theoretical Astrophysics, University of Oslo, Oslo, Norway}

\author{Julien Devriendt}
\affiliation{Dept of Physics, University of Oxford, Keble Road, Oxford OX1 3RH, UK}

\author{Brent Groves}
\affiliation{International Centre for Radio Astronomy Research (ICRAR), University of Western Australia, Crawley, WA 6009, Australia}
\affiliation{Research School of Astronomy \& Astrophysics, Australian National University, Weston Creek, ACT, 2611, Australia}


\author{Matt S. Owers}
\affiliation{Department of Physics and Astronomy, Macquarie University, NSW 2109, Australia}
\affiliation{Astronomy, Astrophysics and Astrophotonics Research Centre, Macquarie University, Sydney, NSW 2109, Australia}

\author{Samuel N. Richards}
\affiliation{SOFIA Science Center, USRA, NASA Ames Research Center, Building N232, M/S 232-12, P.O. Box 1, Moffett Field, CA 94035-0001, USA}

\author{Jesse van de Sande}
\affiliation{Sydney Institute for Astronomy (SIfA), School of Physics, The University of Sydney, NSW 2006, Australia}
\affiliation{ARC Centre of Excellence for All Sky Astrophysics in 3 Dimensions (ASTRO 3D)}

\author{Sarah M. Sweet}  
\affiliation{ARC Centre of Excellence for All Sky Astrophysics in 3 Dimensions (ASTRO 3D)}
\affiliation{Centre for Astrophysics and Supercomputing, Swinburne University of Technology, PO Box 218, Hawthorn, VIC 3122, Australia}




\begin{abstract}

Recent integral field spectroscopy observations have found that about 11\% of galaxies show star-gas misalignment. The misalignment possibly results from external effects such as gas accretion, interaction with other objects, and other environmental effects, hence providing clues to these effects. We explore the properties of misaligned galaxies using Horizon-AGN, a large-volume cosmological simulation, and compare the result with the result of the Sydney-AAO Multi-object integral field spectrograph (SAMI) Galaxy Survey. Horizon-AGN can match the overall misalignment fraction and reproduces the distribution of misalignment angles found by observations surprisingly closely. The misalignment fraction is found to be highly correlated with galaxy morphology both in observations and in the simulation: early-type galaxies are substantially more frequently misaligned than late-type galaxies.
The gas fraction is another important factor associated with misalignment in the sense that misalignment increases with decreasing gas fraction.
However, there is a significant discrepancy between the SAMI and Horizon-AGN data in the misalignment fraction for the galaxies in dense (cluster) environments.
We discuss possible origins of misalignment and disagreement.

\end{abstract}






\section{Introduction}

Stars and gas, the main constituents of galaxies, are closely linked: gas turns into stars, and stars release gas through mass loss. Since angular momentum is conserved during the mass exchange, the rotational axes of stars and gas in a galaxy are expected to be aligned. However, earlier observations have found that some galaxies have highly misaligned rotations between stars and gas \citep[e.g.,][]{1975PASP...87..965U, 1992ApJ...394L...9R}. Moreover, long-slit observations have revealed that galaxies can be misaligned regardless of their mass or morphology \citep[e.g.,][]{1992ApJ...401L..79B, 1996MNRAS.283..543K, 2001AJ....121..140K, 2016MNRAS.455.2508S}. 
Recently, the advent of integral-field spectroscopy (IFS) observations has revealed more detailed kinematic properties of misaligned galaxies \citep[e.g.,][]{2006MNRAS.366.1151S, 2011MNRAS.412L.113C, 2015A&A...581A..65C, Davis+11, Serra+14, 2014A&A...568A..70B, 2015A&A...582A..21B, 2015MNRAS.452....2K, 2016MNRAS.461.2068K, Jin+16, 2019MNRAS.483..458B}. The IFS surveys presented the fraction of misaligned galaxies based on their large samples. For example, \cite{2019MNRAS.483..458B} reported that about 11\% of observed galaxies are misaligned, with position angle offsets between the stellar and gas rotational axes being larger than 30 degrees.

There have been many observational and theoretical studies aiming to reveal the origins of star-gas misalignment. Observations suggest that misalignment can be formed if a galaxy accretes gas from a neighboring galaxy or from large-scale filaments 
in misaligned fashions \citep[e.g.,][]{1992ApJ...401L..79B, 2004A&A...424..447P, 2006MNRAS.370.1565C, 2006MNRAS.366..182B}. 
Numerical simulation studies also found hints for origins in the following aspects: (i) galaxy mergers \citep[e.g.,][]{1990ApJ...361..381B, 1991Natur.354..210H, 1996ApJ...471..115B, 1998ApJ...499..635B, 2001Ap&SS.276..909P, 2009MNRAS.393.1255C}, (ii) continuous or episodic gas accretions \citep[e.g.,][]{1996ApJ...461...55T, 2003A&A...401..817B, 2008ApJ...689..678B, 2013MNRAS.428.1055A, 2014MNRAS.437.3596A, 2015MNRAS.451.3269V}, and (iii) interactions with nearby galaxies \citep[e.g.,][]{2004A&A...426...53D}. 
These simulations were based on a small number of galaxies or ``idealized'' cases. Considering that misalignment is a highly non-linear phenomenon, we need more comprehensive research based on data with a statistically meaningful size.

Large-volume cosmological simulations offer advantages for studying the star-gas misalignment. Within the simulations, various galaxies with a wide variety of masses, morphologies, and environments evolve in the cosmological context. The large number of galaxies allows us to take a statistical approach. For example, we can investigate how these parameters affect misalignment. Moreover, we can simultaneously observe the past, the present, and the future of misaligned galaxies to identify the sequence of formation and the evolution of misalignment. 

\cite{2019ApJ...878..143S} recently performed an important investigation of this issue based on the Illustris simulation and found that (i) Supermassive black hole feedback and gas stripping during fly-by passages through group environments are the two main channels of misalignment, (ii) several galaxies maintain misaligned components for more than 2 Gyr, and (iii) early-type or gas-poor galaxies have higher misaligned fractions. Our study confirms some of their key results and presents additional results based on a different simulation, as described in detail in the following sections.

We investigated misaligned galaxies using the large-volume Horizon-AGN simulation \citep{2014MNRAS.444.1453D}. Here in this paper (Paper I), we examine the properties of misaligned galaxies using the simulation and compare them with the data from the Sydney-AAO Multi-object Integral field spectrograph (SAMI) Galaxy Survey \citep{2015MNRAS.447.2857B, 2012MNRAS.421..872C}. 
In this Paper I, we first try to check how simulations compare with observations in terms of the overall misalignment fraction and the distribution of misalignment angles. Then, we move on to the trend of the misalignment fraction as a function of galaxy morphology, mass, gas fraction, and environment. 
In the following Paper II, we will investigate the formation channels of the misaligned galaxies and quantify the significance of each channel. We will also examine the survival timescale of the star-gas misalignment depending on the properties of the host galaxy. Ultimately, our goal is to identify how gas flows into galaxies and how gas accretion affects galactic evolution.

This paper is organized as follows. In Section~\ref{sec:2}, we will describe how we select galaxies for the Horizon-AGN and SAMI data. In Section~\ref{sec:3}, we will examine the misaligned galaxies in Horizon-AGN focusing on how the misalignment fraction changes depending on the properties of the galaxies. Also, we will compare the misaligned galaxies in the observations and the simulation. Finally, we will discuss our results in Section~\ref{sec:4}. We will also examine the impact of the ``grid-locking effect'' on our study in Appendix \ref{sec:AppA}.

\section{Methodology}
\label{sec:2}
\subsection{The Horizon-AGN simulation}
\label{sec:hagn} 

Horizon-AGN \citep{2014MNRAS.444.1453D} is one of the state-of-the-art hydrodynamical simulations, run with the AMR code {\sc{ramses}} \citep{2002A&A...385..337T}, within the cosmological context from the seven-year Wilkinson Microwave Anisotropy Probe results \citep{2011ApJS..192...18K}. The side length of the simulation box is 100 Mpc/h, and the maximum (smallest) force resolution is about 1 kpc. The mass resolution is $8\times10^7 M_{\sun}$ for dark matter, and $2\times10^6 M_{\sun}$ for stellar particles. Horizon-AGN has 787 snapshots with a time interval of about 17 Myr, but these snapshots have stellar particles only. Sixty-one out of 787 snapshots have full data including stars, gas, dark matter, and sink (black hole) particles. Their time interval is about 250 Myr. Readers are referred to \citet{2014MNRAS.444.1453D} for more details. 

\subsection{Galaxy identification}
\label{sec:hagn_gal} 
Galaxies in the simulation were identified using HaloMaker through the AdaptaHOP algorithm \citep{2004MNRAS.352..376A}, with the most massive sub-node mode \citep{2009A&A...506..647T} applied for stellar particles. A minimum of 50 stellar particles, or $1.7\times10^8 M_{\sun}$, were used to define a galaxy. In Horizon-AGN, we identified 126,362 galaxies at $z=0.055$.
\footnote{We considered the 761st snapshot ($z=0.055$) rather than the last snapshot at $z=0$, for comparison with the observation in this study, mainly because the observed galaxies used in this study are at that distance as well.} 

Galactic models with a small number of stellar particles are not adequate for studying the structure and kinematics of galaxies, as has been discussed in many previous studies.
For example, \cite{2016MNRAS.463.3948D} classified elliptical galaxies from Horizon-AGN with $V/{\sigma} \leq 1$.
They investigated the fraction of elliptical galaxies as a function of galaxy stellar mass and found good agreement with observations when $M_* \gtrsim 2 \times 10^{10} M_{\sun}$. 
Similar exercises and conclusions have been made based on other simulations \citep{2015MNRAS.454.1886S, 2017MNRAS.467.3083R, 2017MNRAS.468.3883P}.
Therefore, we limit our study to galaxies with stellar mass above $10^{10}M_{\sun}$ which corresponds to $\sim 3,000$ stellar particles. 
In the case of Horizon-AGN, the number of galaxies with stellar mass above $10^{10}M_{\sun}$ is 27,908 out of 126,362 at $z=0.055$. 

We use kinematic classification of morphology. 
Higher $V/{\sigma}$ galaxies tend to have disk-shaped structures due to the highly aligned motion of stars, while lower $V/{\sigma}$ galaxies tend to have spheroidal shape structures. 
The galaxies are classified into early-type galaxies (ETGs) and late-type galaxies (LTGs) using a $V/{\sigma}$ cut of 1 \citep[see][]{2016MNRAS.463.3948D}. 
It should be noted that $V/{\sigma}$ can be measured to be different when measured using different techniques and that the $V/{\sigma}$ of the Horizon-AGN galaxies through mock-IFU measurements show a substantial offset from the values of the observed galaxies \citep{2019MNRAS.484..869V}. The offset, however, does not affect our analysis and conclusion because our analysis is reasonably insensitive to the choice of $V/{\sigma}$ cut, as will be discussed in Section~\ref{sec:morpdistrib}.

\begin{figure}
	\includegraphics[width=\columnwidth]{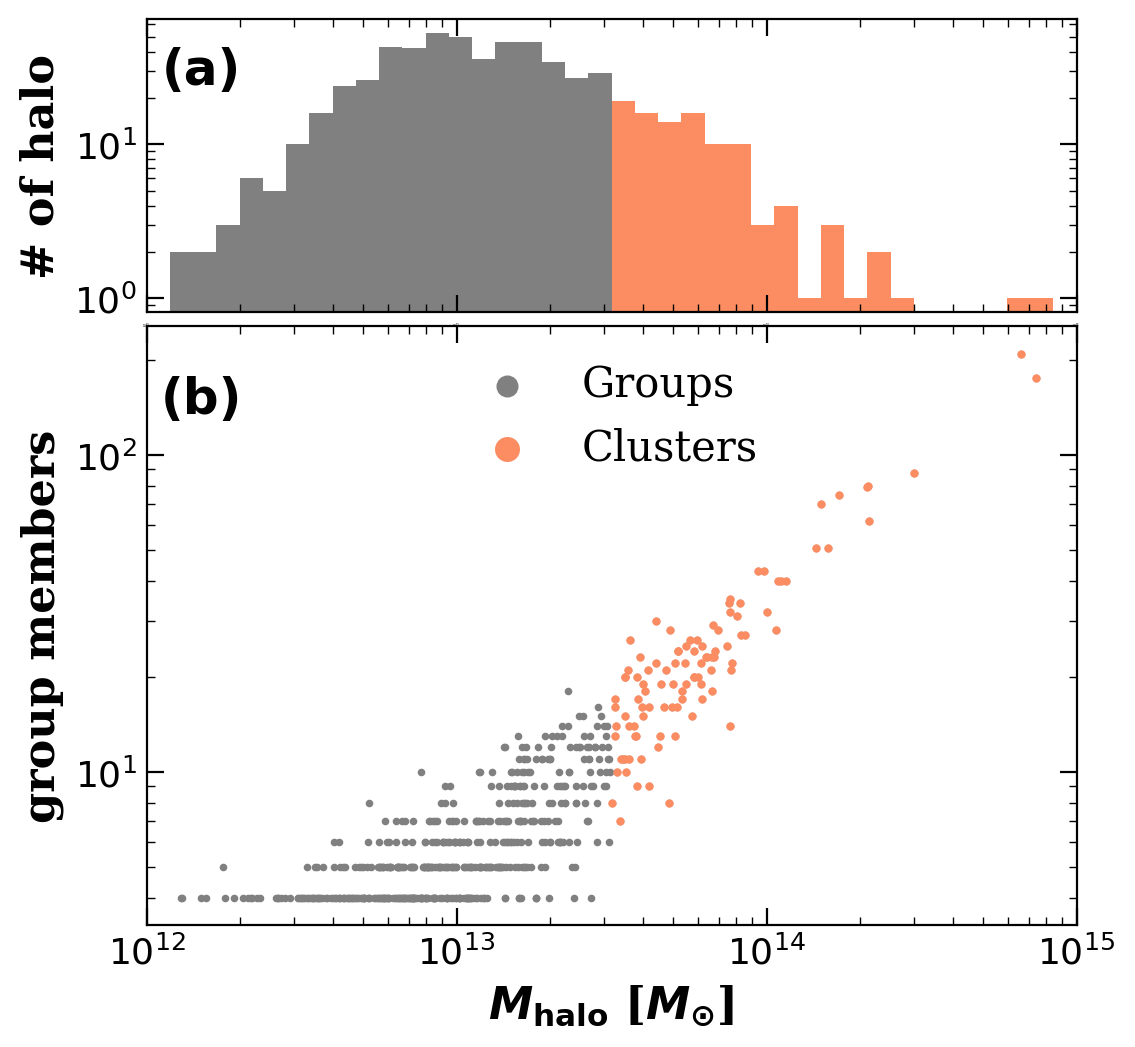}
    \caption{The Horizon-AGN galaxy groups (gray) and clusters (orange). A dark matter halo with a mass range of $10^{12} < M_{\rm vir}/M_{\sun} \leq 10^{13.5}$ having more than 3 members is defined as a group. We define halos with dark matter contents above $10^{13.5}M_{\sun}$ as clusters. Their members are defined when galaxies reside inside 1.5 virial radii ($R_{\rm 200}$). There are 500 groups and 102 clusters in Horizon-AGN. Note that \protect\cite{2019MNRAS.483..458B} used 8 clusters which have virial mass heavier than $10^{14.25}M_{\sun}$.
    \textit{Panel (a)}: the halo mass histogram of groups and clusters. \textit{Panel (b)}: the distribution of the number of member galaxies and the dark matter halo mass.
    }
    \label{fig:image_HAGN_group}
\end{figure}

\begin{figure}
	\includegraphics[width=\columnwidth]{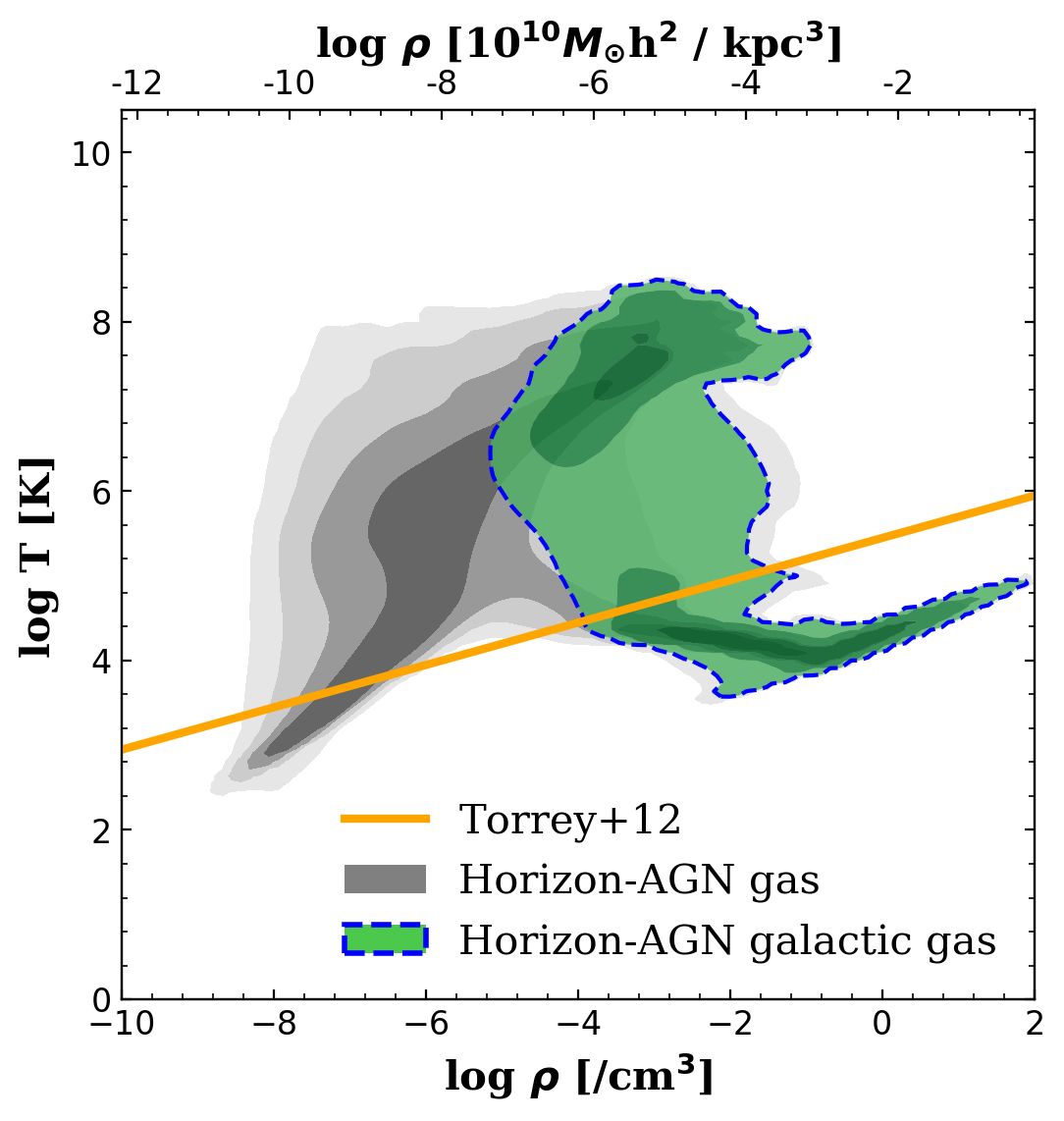}
    \caption{The temperature-density diagram of Horizon-AGN gas cells. We mark the 1, 2, 3, and 4 sigma contours for the whole Horizon-AGN gas cells (gray) and 1, 2, and 3 sigma contours for the cells inside $1\,R_{\rm eff}$ of all the galaxies (green). The yellow line shows the demarcation between hot and cold gases from \protect\cite{2012MNRAS.427.2224T}.}
    \label{fig:image_gas}
\end{figure}

\begin{figure*}
	\includegraphics[width=\linewidth]{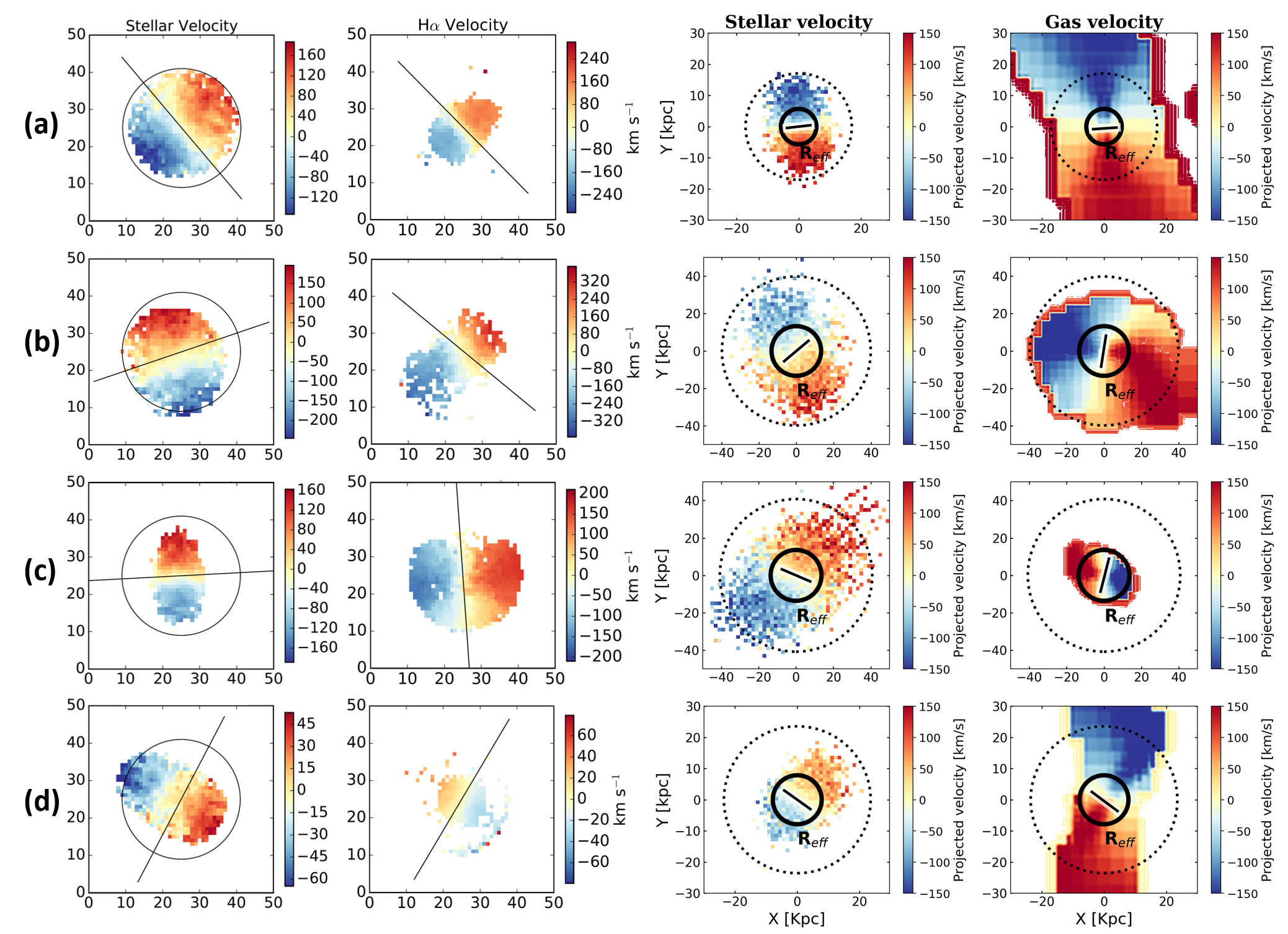}
    \caption{Star and gas velocity maps of example galaxies from SAMI \protect\citep[][the two left columns]{2019MNRAS.483..458B} and Horizon-AGN (the two right columns). We used ``cold gas'' instead of ionized gas in the simulation.
    From top to bottom, four types of galaxies, (a) aligned, (b) misaligned, (c) polar disk (PA offset $\sim 90 ^{\circ}$), and (d) counter-rotating galaxies (PA offset $\sim 180 ^{\circ}$), are demonstrated. Each rotational axis is expressed by a black line.
    The circle in the SAMI plot indicates the 15 arcsec size of the SAMI hexabundle. The black circles in the two right columns show $1\,R_{\rm eff}$ (solid) and $3\,R_{\rm eff}$ (dotted) of the galaxy.}
    \label{fig:image_IFU}
\end{figure*}

\subsection{Cluster and group identification}
\label{sec:hagn_clus}

We identified groups and clusters in Horizon-AGN as follows. We first identified dark matter halos with $M_{\rm vir} > 10^{11}M_{\sun}$ and counted their member galaxies with $M_* > 10^{10}M_{\sun}$ within 1.5 virial radii ($R_{\rm 200}$). 
A halo is classified as a ``group'' when its mass is in the range of $10^{12} < M_{\rm vir}/M_{\sun} \leq 10^{13.5}$ and its member galaxies number at least 4.
Clusters are defined as having greater mass than that ($M_{\rm vir}/M_{\sun} > 10^{13.5}$).
Thus, Horizon-AGN is found to contain 500 groups and 102 clusters at z=0.055.
Fig.~\ref{fig:image_HAGN_group}-(a) shows the histogram of Horizon-AGN groups (gray) and clusters (orange). 
We note that Horizon-AGN and typical volume simulations with 100 Mpc side-length (e.g., Illustris, Eagle, and so on) would not have so many massive clusters as in the observation.
Also, the mass range of the Horizon-AGN clusters is intentionally kept broad so that we can investigate the possible presence of cluster (halo) mass dependence on the misalignment.
Fig.~\ref{fig:image_HAGN_group}-(b) shows the number of member galaxies against the group mass. 
The total numbers of member galaxies are 5,924 and 2,711 inside groups and clusters, respectively. 

\subsection{Galactic gas and the rotational axis}
\label{sec:hagn_gas}

Fig.~\ref{fig:image_gas} shows the temperature-density phase diagram of all the Horizon-AGN gas cells (grey) and the galactic gas (green). 
We define the galactic gas that the gas cells inside one effective radius ($R_{\rm eff}$), encompassing half of the total stellar mass (half projected stellar mass) of the galaxy.
Since we are interested in the gas properties of individual galaxies, we divide the galactic gas in the simulation into galactic cold gas and (non-cold) ``surroundings'' as follows. 
We use a linear cut in the logarithmic density-temperature plane using Equation~(\ref{eq:torrey}) from \cite{2012MNRAS.427.2224T}:
\begin{equation}
    \log(T/[K])=6+0.25\log(\rho/10^{10} [M_\odot h^2 \rm{kpc} ^{-3}]).
	\label{eq:torrey}
\end{equation}
``Galactic cold gas'', the low-temperature side, is used for star formation in the simulation. 
According to this scheme, it has a temperature of roughly 10,000-30,000 K depending on the density. 
The cold gas is located at the central and disk parts of the galaxy and found to represent the kinematic property of the interstellar medium.
Also, the motion of the cold gas will correspond to the observed gas motion, because IFS observations such as SAMI measured gas motion using H$\alpha$ emission lines ($\sim 10,000$ K). We note that Horizon-AGN simulation cannot resolve the multi-phase nature of the ISM, including molecular gas. The ``cold gas'' here does not literally mean the cold gas as found by observations.
``The surrounding gas'', on the other hand, is the gas above the density-temperature criterion. 
It corresponds to intracluster medium or intergalactic medium.  
To investigate the rotation of gas in the galaxies, we focus only upon the cold gas.

To measure the rotational axes of a galaxy, we measure the velocity of all stellar particles and gas cells belonging to the galaxy. We measure their angular momentum with their position and velocity relative to the galactic center. We define the rotational axes of the galaxy in terms of the direction of the net angular momentum of each component. 
We measure the rotational axes inside one effective radius and measure the position angle (PA) offset between the rotational axes of the stars and gas. The PA offset (misaligned angle) is defined to be within the range of 0 (aligned) to 180 degrees (counter-rotating). 

We draw stellar and gas velocity maps similar to IFS data using Horizon-AGN galaxies. Some examples are shown in Fig.~\ref{fig:image_IFU}, with the observational data from SAMI \citep{2019MNRAS.483..458B} for comparison. The effective radius is marked with a black solid circle, and each rotation axis is shown with a black line. The simulation has reproduced many different types of misaligned galaxies, including polar disk galaxies (PA offset $\sim 90 ^{\circ}$) and counter-rotating galaxies (PA offset $\sim 180 ^{\circ}$). 

We note that the directions of the rotation of stars and the gas in the grid-based simulation are more likely to be aligned with the direction of the grid. \textbf{Since it is affected by the grids, it is possible that our analysis is likely affected by the grid resolutions, too. The Horizon-AGN simulation is available only for one AMR resolution setting, and so we have not performed a resolution test in this analysis.} We present the impact of the so-called ``grid-locking'' effect \citep{2015MNRAS.454.2736C, 2020MNRAS.tmp..237K} on our analysis in Appendix \ref{sec:AppA}.

\section{Result}
\label{sec:3}

\subsection{Comparing with the SAMI data}
\label{sec:result_IFU} 

\subsubsection{SAMI samples}
\cite{2019MNRAS.483..458B} used the SAMI sample for which PAs are measured for both gas and stars: 486 out of 833 galaxies in the field/group regions and 136 out of 380 in cluster regions. They measured the PA offset from the difference between the two PAs (stars and gas). According to \cite{2019MNRAS.483..458B}, the error of the fitted PAs is estimated to be $\pm10$ degrees and to be much larger for galaxies with a low gas content or with low signal-to-noise-ratio spectrum. They classified galaxies as ``misaligned'' when they had a PA offset larger than 30 degrees for direct comparison to the previous papers \citep[e.g.,][]{2015MNRAS.448.1271L, Davis+16}. In this paper, we also used their criterion of star-gas misalignment to compare our results.

SAMI classified the morphology of galaxies by visual inspection. The detailed method can be found in \cite{2016MNRAS.463..170C}. The masses of the field/group galaxies range from $10^8 M_{\sun}$ to $5\times10^{11} M_{\sun}$, while those of the cluster sample range $10^{10} M_{\sun}$ through $5\times10^{12} M_{\sun}$. 
We would like to remind the readers that the lower mass limit of our Horizon-AGN galaxies is $10^{10} M_{\sun}$.
A substantial fraction of SAMI sample galaxies are below this mass cut. We have compared the mass distributions of two samples for the same mass range (above $10^{10} M_{\sun}$), and they are reasonably close to each other. If we use the subsample of SAMI galaxies above $10^{10} M_{\sun}$ alone, the misalignment fraction changes only slightly (to 12.9\% field, 11.2\% cluster). Given many other uncertainties, we thought this difference was too small to require a different sampling strategy.

The virial masses of the SAMI clusters are $10^{14.25} < M_{\rm 200}/M_{\sun} < 10^{15.19}$ \citep{2017MNRAS.468.1824O}, which is much greater than those of the Horizon-AGN clusters shown in Fig.~\ref{fig:image_HAGN_group}. We will discuss the impact of this difference in Section~\ref{sec:diff_obs_sim}.
The redshift range of the SAMI data is up to 0.1, but its cluster galaxies have a narrower redshift range ($0.02<z<0.07$). The detailed description of the SAMI cluster can be found in \cite{2017MNRAS.468.1824O}.

\begin{figure}
	\includegraphics[width=\columnwidth]{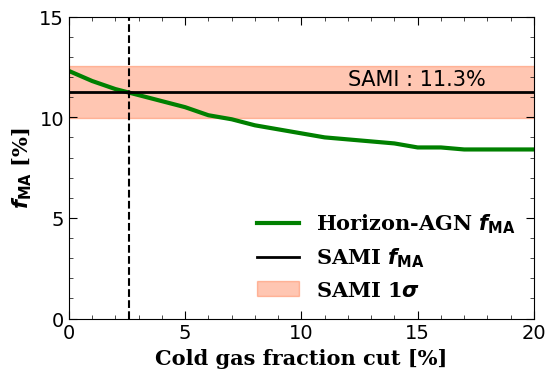}
    \caption{The Horizon-AGN misalignment fraction depending on the cold gas fraction cut. We regard the cold gas fraction criterion as a gas detection limit. We measure the misalignment fraction (green line) of ``observable galaxies'' which are above a certain gas fraction. While they show a monotonic, negative trend, Horizon-AGN galaxies with $f_{\rm gas} \gtrsim 0.03$ (vertical dotted line) reproduce the SAMI misalignment fraction of 11.3\% (horizontal line). Note that the misalignment fraction in the simulation is not very sensitive to the gas fraction cut. The binomial $1 \sigma$ error of the SAMI sample is expressed in the shaded area.}
    \label{fig:image_gfrac}
\end{figure}

\begin{figure*}
	\includegraphics[width=\linewidth]{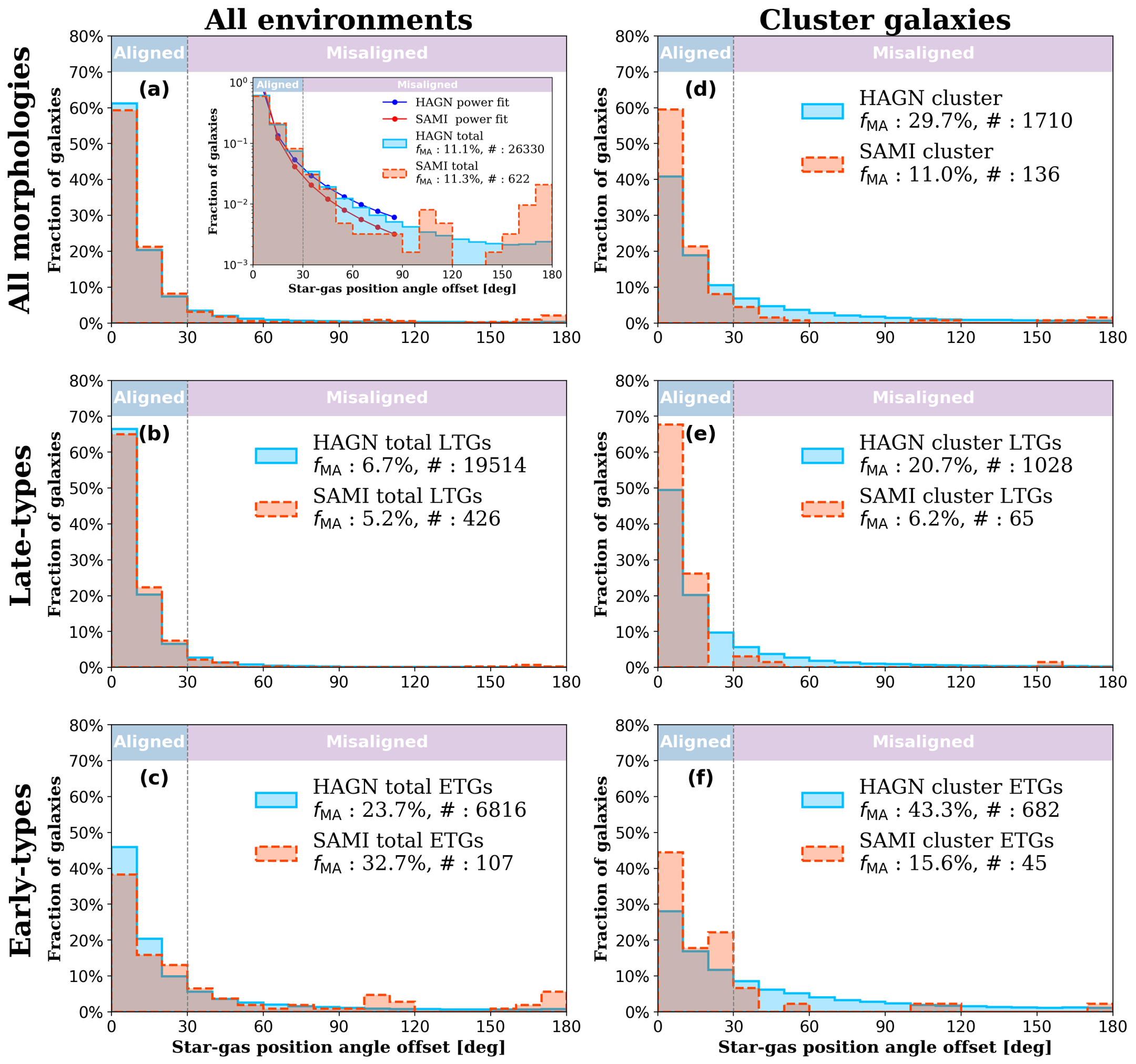}
    \caption{The distribution of star-gas PA offset in Horizon-AGN (blue) and SAMI (red) galaxies. We have performed a Monte-Carlo simulation of projection on the Horizon-AGN galaxies as many as 1,000 times to minimize bias from the projection effect. We classified misaligned galaxies when their PA offset exceeds 30 degrees, following previous researches \protect\cite[e.g.,][]{2019MNRAS.483..458B}. The legend shows the misalignment fractions and the number of samples. The left column shows all galaxies, regardless of their environment, while the right column shows galaxies belonging to clusters. \textit{Top panels}: the galaxies from Horizon-AGN and SAMI, regardless of their morphological classification. For Panel (a), the inset diagram presents the same histograms with the best-fit power laws in a logarithmic scale.
    \textit{Middle panels}: the distribution of late-type galaxies ($V/\sigma > 1$). \textit{Bottom panels}: the distribution of early-type galaxies ($V/\sigma \leq 1$).}
    \label{fig:image_SAMI_dist}
\end{figure*}

\subsubsection{Gas detection limit}
\label{sec:samize:gas}

The SAMI misalignment study \citep{2019MNRAS.483..458B} used only the galaxies whose PAs have been measured for both gas and stars. A considerable number of observed galaxies were excluded due to the low gas emission flux. 
Therefore, it is necessary to set a detection limit for gas kinematics in the simulation that is consistent with observations.

The SAMI study \citep{2019MNRAS.483..458B} does not have comparable measures of the gas contents or gas fractions of its galaxies.
Thus, we determine the gas detection limit indirectly from the misalignment fraction, which depends upon the cold gas fraction.
We define the gas fraction as the mass ratio of cold gas to stars ($M_{\rm cold\,gas}/M_{\rm *}$) within $1\,R_{\rm eff}$ of the galaxy.
After that, we regard the gas fraction as a gas detection limit; more gas-rich galaxies are considered better ``observable''. We argue that the gas fraction is more important than the gas mass in determining the detection limit because the SAMI team measured the gas kinematics using H$\alpha$ emission. They took spectrum, which is the mixture of light coming from both stars and gas, and subtracted the stellar continuum from it using PPXF \citep{2004PASP..116..138C}. In this procedure, the stellar light acts as a noise. Therefore, the mass ratio of gas to stars is more adequate as a proxy of gas detection limit.
Fig.~\ref{fig:image_gfrac} shows a monotonic, negative trend between the misalignment fraction and the gas fraction cut. 
The Horizon-AGN galaxies with $f_{\rm gas} > 0.03$ reproduce the SAMI misalignment fraction of 11.3\%, which is the fraction of galaxies with a PA offset exceeding 30 degrees.
We, therefore, conclude that the gas detection limit roughly corresponds to a gas fraction of 3\% within $1\,R_{\rm eff}$. 
With this criterion, the total number of ``observable'' galaxies is 26,330 at $z=0.055$. 
Although this is an arbitrary value, we want to note that the misalignment fraction in the simulation is not very sensitive to the gas fraction according to Fig.~\ref{fig:image_gfrac}.

\subsubsection{The distribution of the star-gas PA offset}
\label{sec:distrib} 
Observational data are projected to the perpendicular plane to the line-of-sight. 
In order to minimize bias from the projection effect, we have performed a Monte-Carlo simulation of projection as many as 1,000 times on each model galaxy and measured their PA offsets. 

The histogram of star-gas PA offset is shown in Fig.~\ref{fig:image_SAMI_dist}. Panel (a) shows the histograms of the 622 galaxies from SAMI (red) and the 26,330 (times 1,000 projections) galaxies from Horizon-AGN (blue). \cite{2019MNRAS.483..458B} reported that about 11.3 $\pm$ 1.2\% of galaxies are misaligned ($\rm PA \ offset > 30^{\circ}$), and Horizon-AGN galaxies with $f_{\rm gas} > 0.03$ show a misalignment fraction of 11.1 $\pm$ 0.1\% \footnote{The errors are simple binomial errors and do not include other systematic errors.}. 
The close agreement between SAMI and Horizon-AGN \textbf{shown} here is simply a result of our calibration using the gas fraction. In addition, one should also be aware of systematic uncertainty, for example, due to the grid-locking effect (see Appendix \ref{sec:AppA}), in the measurement of the directions of stars and the gas in simulations. \textbf{As mentioned earlier, grid resolutions likely affect the measurement of misalignment. In this analysis we have not performed a resolution test.}

The Horizon-AGN galaxies show a similar distribution to that of the SAMI galaxies. 
We evaluate the likelihood that the two distributions are drawn from the same population, using a two-sample Kolmogorov-Smirnov (KS) test. The KS-test statistic is 0.034 and the p-value is 0.454, which suggests that they are likely to be drawn from the same population.

We performed a power-law fit to the histogram in the log-log plane to guide the eye in comparing the distribution from SAMI with the simulations, but note that there is not a physical basis for this fit. The fit was performed for the range of 0--90 degrees of offset angle only.
The best-fit power-law indices are $-2.1$ for SAMI and $-1.8$ for the Horizon-AGN data. 

We also present the same histogram with a logarithmic scale in the inset diagram to highlight the apparent two peaks at high PA offsets. 
Prominent differences between the SAMI and Horizon-AGN data appear around 90 and 180 degrees.
\cite{2019MNRAS.483..458B} suggested that those peaks are linked with the dynamical settling-down processes of PA offsets. 
The stability of the gas disk depends on the loss of the angular momentum of the gas disk. The star-gas position angle offsets at 0 and 180 degrees are stable states. They will maintain their offset until the gas is consumed by star formation \citep{2017MNRAS.471L..87O}. The star-gas misalignment at 90 degrees is a semi-stable state that the gas disk stays longer than other orbits (see Section~\ref{sec:hagn_morp}).

Similar peaks were found in the ATLAS 3D dataset \citep{Davis+16}. Also, the polar disk structures of S0 galaxies \citep[e.g.,][]{1983AJ.....88..909S, 1987ApJ...314..457V, 1990AJ....100.1489W} imply the presence of a 90-degree peak. 
Horizon-AGN, however, does not reproduce the 90- and 180-degree peaks. 
This might be due to the insufficient force-calculation resolution of the simulation, and the stability of the two peaks may depend on how accurately thin disks are realized in the simulation.
Horizon-AGN has a maximum (best) force-calculation resolutions of roughly 1 kpc, and with this, it is difficult to resolve/reproduce a thin disk. 
Therefore, in Horizon-AGN, galactic gas disks are not as thin as in real galaxies. 
The outer parts of the apparently-thick disk will start to feel an imbalance in torque from competing directions (i.e., from the stellar disk). 
As a result, misalignments of around 90 degrees, for example, may decay more easily when the resolution is insufficient.

\subsubsection{Morphology and the misalignment}
\label{sec:morpdistrib} 

The star-gas PA offset distributions of LTGs and ETGs are shown in Figs.~\ref{fig:image_SAMI_dist}-(b) and -(c), respectively. 
LTGs tend to be more aligned than ETGs in both the observation and the simulation. 
For a reference, \cite{2019ApJ...878..143S} found a consistent result regarding the morphology dependence of the misalignment fraction.
The misalignment fractions of LTGs are comparable between SAMI (5.2 $\pm$ 0.7\%) and Horizon-AGN (6.7 $\pm$ 0.1\%). 
The shapes of the histograms are also consistent between them: KS-test statistic = 0.034, and p-value = 0.691.

On the other hand, the misalignment fractions of ETGs are substantially different between the two samples: 32.7 $\pm$ 6.6\% (SAMI) and 23.7 $\pm$ 0.2\% (Horizon-AGN). 
The overall distribution of the PA offset too is markedly different: KS-test statistic = 0.130, and p-value = 0.050.

The discrepancy between the results from SAMI and Horizon-AGN may come from multiple origins. 
One may be due to the different classification methods applied. 
The Horizon-AGN galaxies are classified as ETGs and LTGs using the cut of $V/{\sigma} = 1$, while the SAMI galaxies are classified via visual inspection. 
For example, a good fraction of S0 galaxies classified as ETGs by visual inspection may be classified as LTGs by the $V/{\sigma}$ criterion.
If we change the morphology criterion to $V/{\sigma} = 0.8$, the misalignment fraction in Horizon-AGN becomes 29.5 $\pm$ 0.3\% (KS-test statistic = 0.078, and p-value = 0.485).


\subsubsection{Cluster environment and misalignment}
\label{sec:envdistrib} 

The star-gas PA offset distribution and misalignment fraction of cluster galaxies are shown in Fig.~\ref{fig:image_SAMI_dist}-(d). Since only a small fraction of Horizon-AGN galaxies belong to the cluster environment ($\sim 10\%$, or less), the histograms of group/field galaxies would be nearly the same as those of all galaxies (the left panels). 
In the case of Horizon-AGN, we find that cluster galaxies have a misalignment fraction 2.7 times higher (29.7 $\pm$ 0.5\%) than the group/field galaxies (11.1 $\pm$ 0.1\%). 
This result significantly disagrees with the SAMI galaxies that show no clear difference is apparent between cluster and non-cluster samples (KS-test statistic = 0.236, and p-value = 3.710e-7).

We note in Figs.~\ref{fig:image_SAMI_dist}-(b) and -(c) that the misalignment fraction is higher for ETGs.
Thus, since ETGs are more frequently found in dense environments \citep{1980ApJ...236..351D}, the misalignment fraction is also expected to be higher in denser environments. 
In this sense, part of the higher misalignment fraction of cluster galaxies in Horizon-AGN can be understood. 
However, this is not the whole story when we divide the cluster galaxies into LTGs (Fig.~\ref{fig:image_SAMI_dist}-(e)), and ETGs (Fig.~\ref{fig:image_SAMI_dist}-(f)). 
The misalignment fraction of the cluster galaxies in Horizon-AGN (29.7\%) is even higher than that of the general ETGs (23.7\%), meaning that even if clusters are entirely made up of ETGs, their misalignment fractions cannot be explained by the morphology mix alone. 

We also found higher misalignment fractions in cluster environments in the Horizon-AGN simulation, regardless of the morphologies of galaxies (LTG: 20.7 $\pm$ 0.5\%, ETG: 43.3 $\pm$ 0.9\%) by a factor of 3.1 (LTG) and 1.8 (ETG) compared to the whole sample (LTG: 6.7\%, ETG: 23.7\%). 
Thus, something other than just the morphology mix is affecting the misalignment of cluster galaxies, perhaps through some environmental effects (see Section~\ref{sec:dis_env}).
\cite{2019MNRAS.483..458B}, however, reported in their SAMI observations that the misalignment fraction of LTGs is increased in the cluster environment, whereas the misalignment fraction of ETGs is greatly reduced.
While the former is consistent with Horizon-AGN, the latter is not.
Meanwhile, both the misalignment fractions and distributions are quite different. 
Their KS-test statistic is 0.248, and p-value is 5.271e-4.
The two ETG samples also have different PA offset distributions (KS-test statistic = 0.302, and p-value = 3.908e-4).
At this stage, we find it difficult to understand the origin of this discrepancy: it could potentially come from the differences in gas measurement methods. We will discuss this phenomenon and the reasons behind it further in Section~\ref{sec:dis_env}.

\begin{figure*}
	\includegraphics[width=\linewidth]{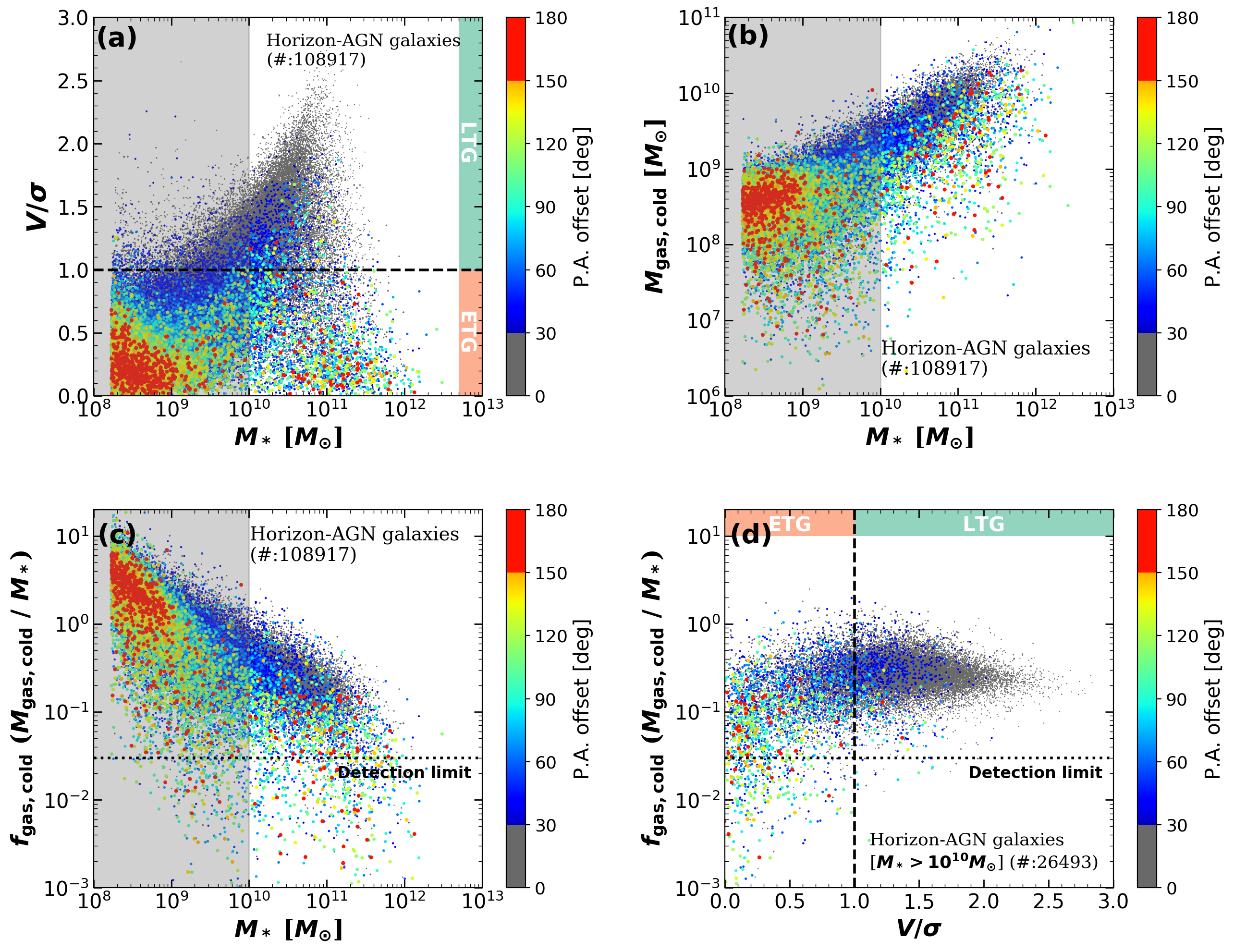}
    \caption{The distribution of the star-gas PA offset depending on the properties (stellar mass, cold gas mass, cold gas fraction and $V/{\sigma}$) of Horizon-AGN galaxies. 
    All points in the panels are color-coded based on the PA offset.
    Each point-size scales with PA offset to highlight misaligned galaxies.
    Low-mass ($M_* < 10^{10} M_{\sun}$) galaxies in the gray shaded regions in panels (a), (b), and (c) have been excluded from the analysis. Panel (d) shows only the galaxies with stellar masses above $10^{10} M_{\sun}$. 
    The cold gas mass and the cold gas fraction are measured inside $1\,R_{\rm eff}$ of the galaxy. The detection limit ($f_{\rm gas} = 0.03$) is expressed as a black dotted line. The black dashed line ($V/{\sigma}=1$) divides galaxies into LTGs and ETGs. Overall, lower $V/{\sigma}$ and gas-poor galaxies are more likely to be misaligned. 
    }
    \label{fig:image_HAGN_phy}
\end{figure*}

\begin{figure*}
	\includegraphics[width=\linewidth]{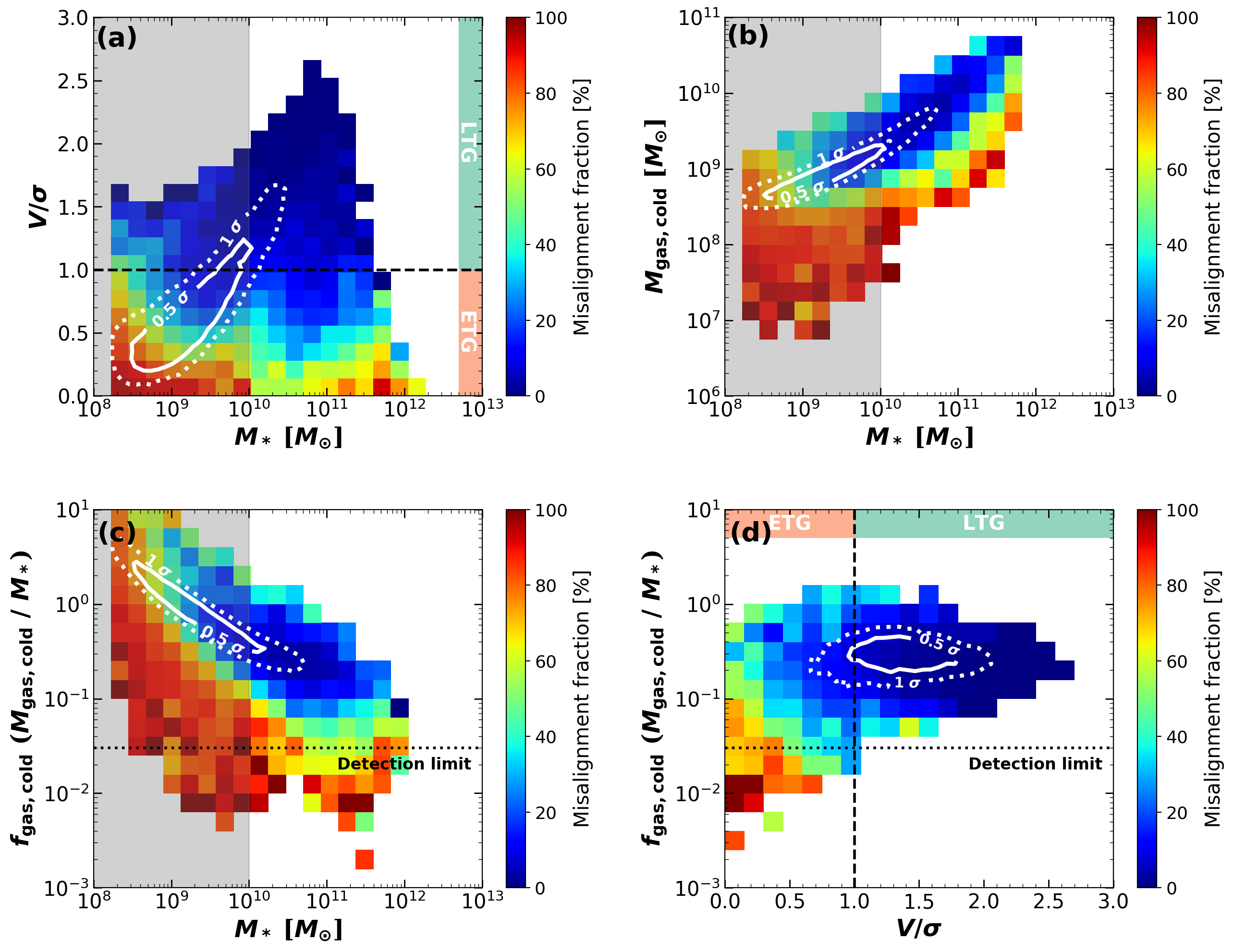}
    \caption{2D histogram showing the star-gas misalignment fraction depending on the properties of Horizon-AGN galaxies based on Fig.~\ref{fig:image_HAGN_phy}. We classified misaligned galaxies when their PA offset exceeds 30 degrees. The white contours show the $0.5 {\sigma}$ (solid line) and $1 {\sigma}$ (dotted line) distributions of galaxies. Each pixel contains at least 5 galaxies to ensure statistical significance.}
    \label{fig:image_HAGN_2Dphy}
\end{figure*}

\begin{figure}
	\includegraphics[width=\columnwidth]{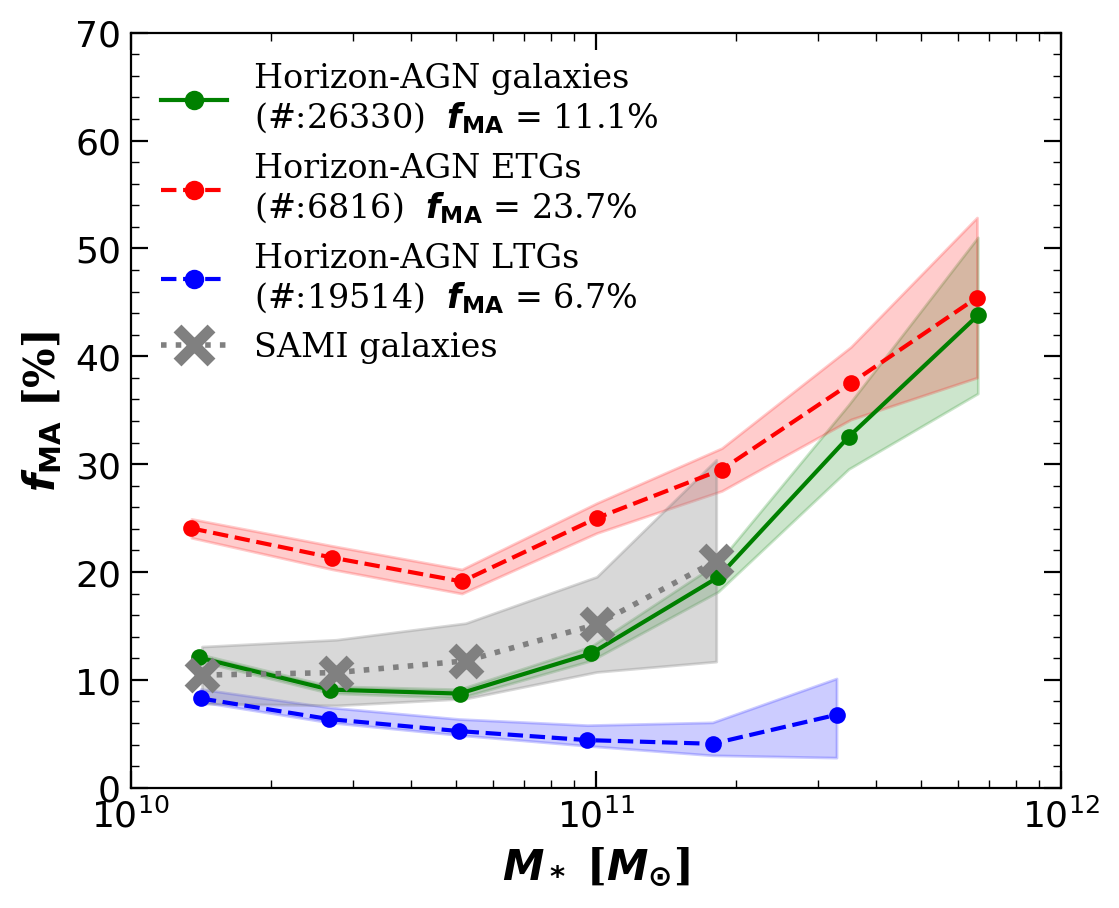}
    \caption{The misalignment fraction of Horizon-AGN galaxies as functions of galaxy stellar mass. The gray dotted line is the misalignment fraction measured by \protect\cite{2019MNRAS.483..458B}. The misalignment fractions for all Horizon-AGN galaxies (green), for ETGs (red), and for LTGs (blue) as functions of stellar mass are shown in the figure. Each point contains at least 10 galaxies, and shadowed regions show the $1 {\sigma}$ error of the mean of a binomial distribution. The whole sample (green line) shows that more massive galaxies have higher misalignment fractions, which appears to originate from the mass-morphology relation. While the misalignment fraction of LTGs (blue) remains almost constant, massive ETGs are found to have an enhanced misalignment fraction, since they tend to have a relatively low gas fraction and low $V/\sigma$ ratio.}
    \label{fig:image_detec}
\end{figure}

\begin{figure*}
	\includegraphics[width=\linewidth]{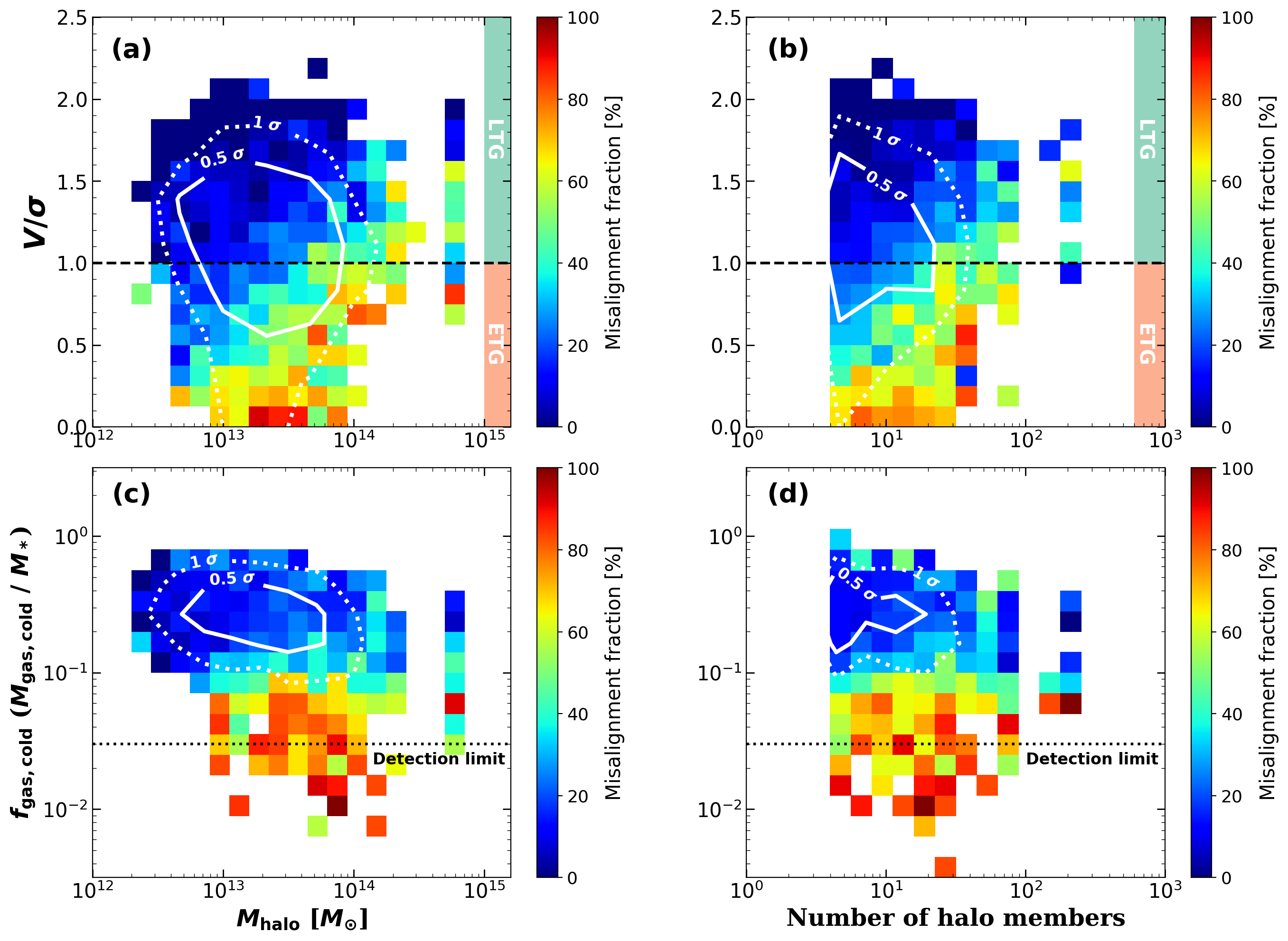}
    \caption{2D histogram showing the misalignment fraction (PA offset $> 30 ^{\circ}$) depending on the group environment of Horizon-AGN galaxies. Each pixel contains at least 5 galaxies. The detection limit ($f_{\rm gas} = 0.03$) is expressed as a black dotted line. The black dashed line ($V/{\sigma}=1$) divides galaxies into LTGs and ETGs. Group and cluster halos with greater masses or number of members have higher misalignment fraction values. 
    \textit{Panel (a)}: the halo mass against $V/{\sigma}$. \textit{Panel (b)}: the number of halo members against $V/{\sigma}$.
    \textit{Panel (c)}: the halo mass against the cold gas fraction. \textit{Panel (d)}: the number of halo members against the cold gas fraction.}

    \label{fig:image_HAGN_2Dgroup}
\end{figure*}

\begin{figure*}
	\includegraphics[width=\linewidth]{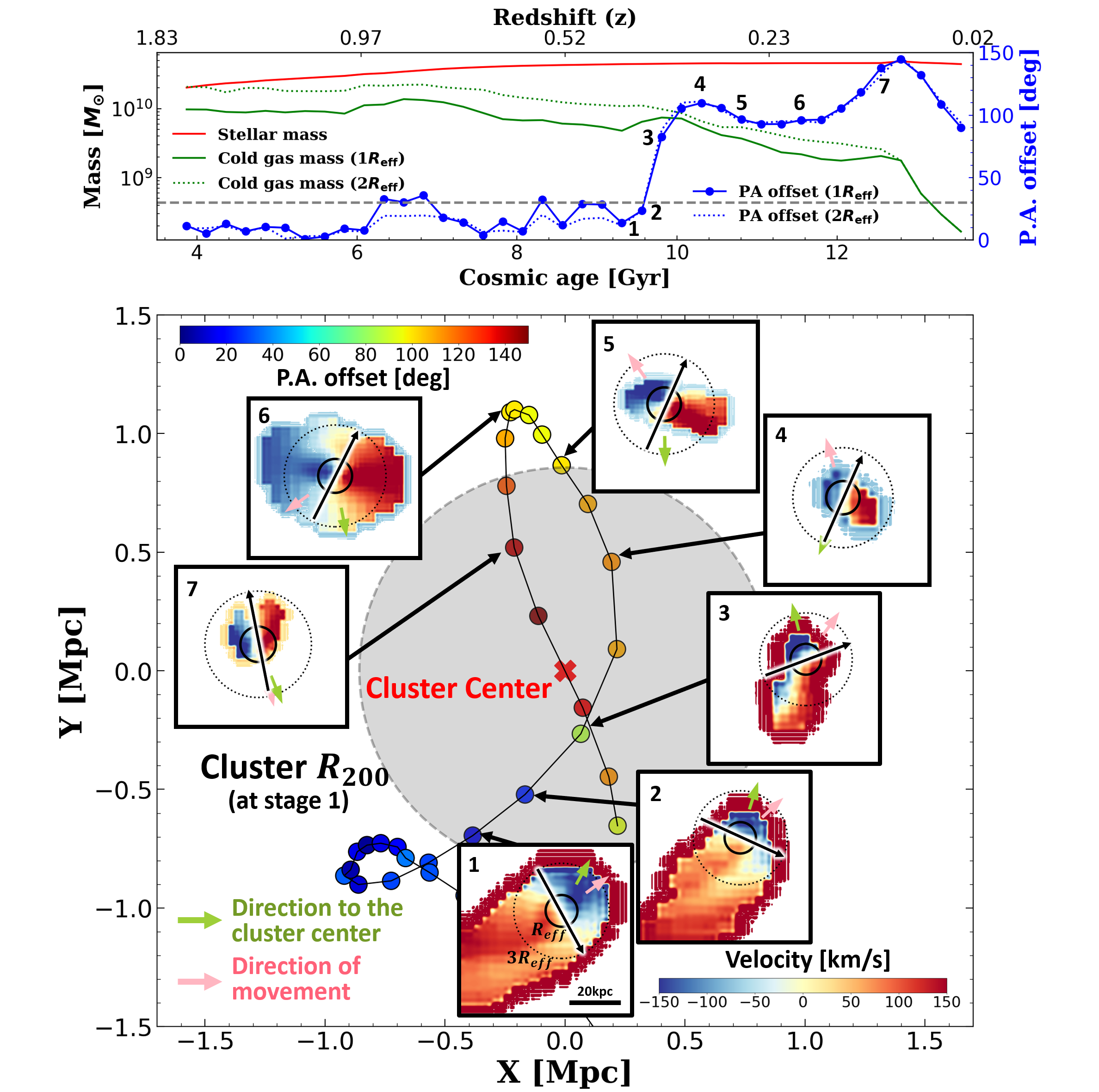}
    \caption{A typical misaligned galaxy in a cluster environment. We numbered particular snapshots along the trajectory (Stages 1 -- 7). \textit{Top}: the properties of the galaxy. 
    The cold gas mass inside 1 and $2\,R_{\rm eff}$ are displayed separately.
    The position angle offsets measured at 1 and $2\,R_{\rm eff}$ are also displayed.
    The horizontal dashed line shows our condition ($> 30$ degree) for misaligned galaxies.
    \textit{Bottom}: the infalling trajectory of the galaxy. The cluster center is fixed and marked with a red X mark. The virial radius ($R_{\rm 200}$) at Stage 1 is expressed by the gray shade region. Each point represents the position of the galaxy, and is color-coded by the PA offset. \textit{Inset panels}: the projected cold gas map of the galaxy. The effective radius (solid circle) and $3\,R_{\rm eff}$ (dotted circle) of the galaxy are shown in the figure. The measured gas rotation axis is expressed in a black arrow. The green and the pink arrows indicate the direction to the cluster center and the galaxy's motion, respectively. While the stripped gas tail extends in the opposite direction to the galaxy's motion (ram pressure), the position angle of a gas disk changes non-systematically (“wobbles”) during the orbital motion of the galaxy inside the cluster.}
    \label{fig:image_clusterdriven}
\end{figure*}

\begin{figure}
	\includegraphics[width=\columnwidth]{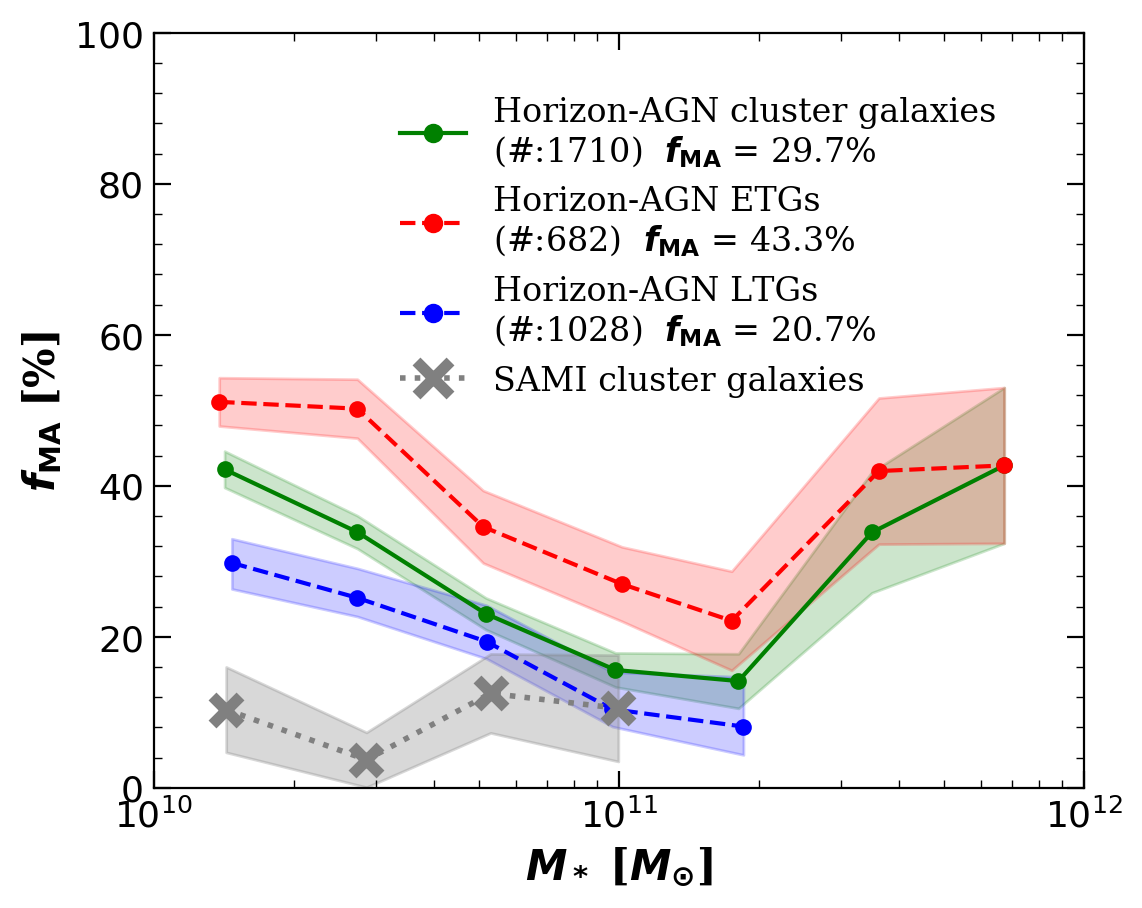}
    \caption{Same as Fig.~\ref{fig:image_detec}, but for cluster galaxies. 
    Each point contains at least 5 galaxies, and shadowed regions show the $1 {\sigma}$ error of the mean of a binomial distribution. While the same mass trend is visible among the cluster galaxies for $M_* \gtrsim 10^{11}$, it shows a reverse mass trend at $M_* \lesssim 10^{11}$ against Fig.~\ref{fig:image_detec}.}
    \label{fig:image_detec_cluster}
\end{figure}

\begin{figure*}
	\includegraphics[width=\linewidth]{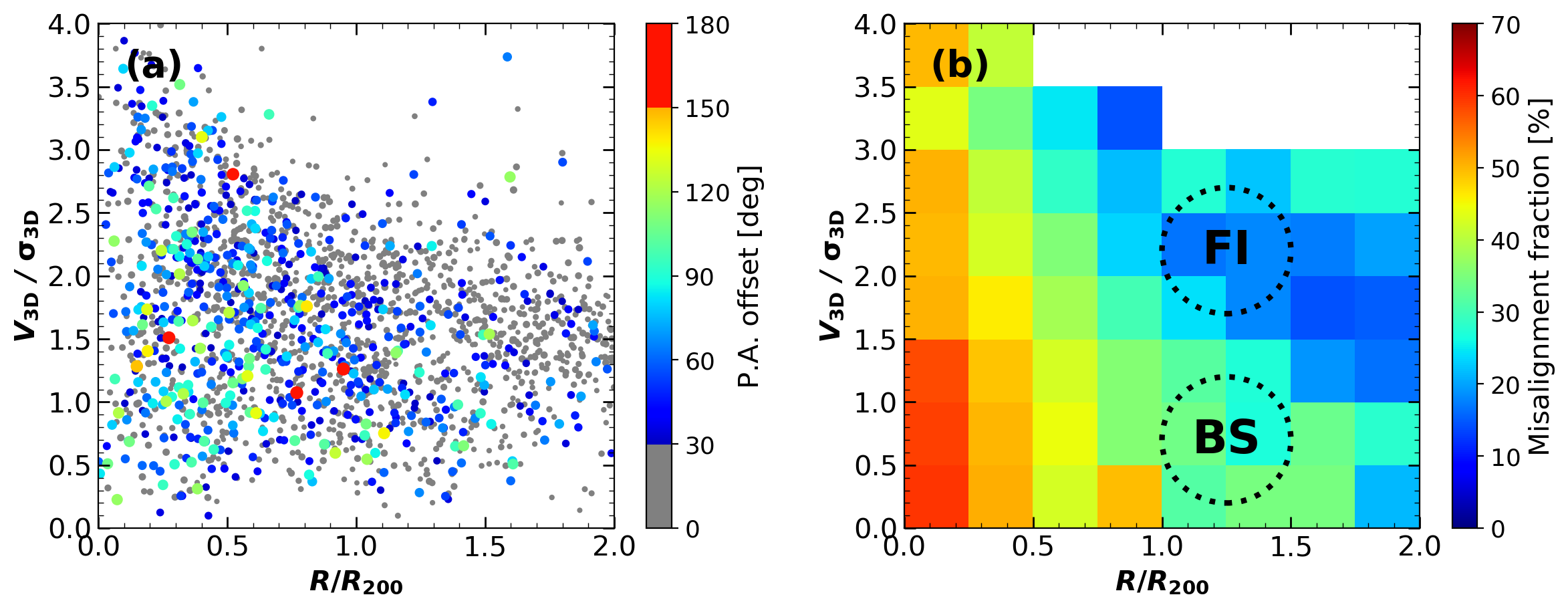}
    \caption{Phase-space analysis of Horizon-AGN cluster galaxies. \textit{Panel (a)}: the phase-space diagram of cluster galaxies. While aligned galaxies are expressed as gray points, the misaligned galaxies (PA offset $> 30^{\circ}$) are color-coded based on their PA offset. Their size scales with their PA offset. \textit{Panel (b)}: 2D histogram of the misalignment fraction of cluster galaxies in the phase-space diagram by stacking the recent 20 snapshots ($z=0 - 0.5$). The central regions have a higher misalignment fraction than those at the outskirts. Each pixel contains at least 20 galaxies. 
    The galaxies that have not fallen into the cluster (``first infaller'') and the ``backsplash'' galaxies are roughly populated in the regions marked as ``FI'' and ``BS'' and associated contours \protect\citep{2017ApJ...843..128R}.
    } 
    
    \label{fig:image_cluster_multi}
\end{figure*}

\subsection{Properties of Horizon-AGN misaligned galaxies}

In this section, we explore the properties of misaligned galaxies, hoping to pin down the main drivers of star-gas misalignment.
Fig.~\ref{fig:image_HAGN_phy} shows the star-gas PA offsets of Horizon-AGN galaxies depending on their properties (i.e., stellar mass, gas mass, gas fraction, and $V/{\sigma}$ ratio). 
Each point is a Horizon-AGN galaxy color-coded based on the star-gas PA offset. The size of each point scales with the PA offset.
Meanwhile, Fig.~\ref{fig:image_HAGN_2Dphy} is a 2D histogram showing the misalignment fraction, which is the fraction of galaxies with a PA offset exceeding 30 degrees, on the same planes of Fig.~\ref{fig:image_HAGN_phy}. 
In the case of Fig.~\ref{fig:image_HAGN_2Dphy}, the 0.5 and $1{\sigma}$ contours are presented. 
Each pixel contains at least 5 galaxies to ensure statistical significance. 
We note that Figs.~\ref{fig:image_HAGN_phy} and \ref{fig:image_HAGN_2Dphy} contain the galaxies with a gas fraction lower than the gas detection limit ($f_{\rm gas} > 0.03$). Here, we focus only on the Horizon-AGN galaxies, and thus the gas detection limit is irrelevant.

In these two figures, we used all of the Horizon-AGN galaxies with $M_* > 2 \times 10^{8} M_{\sun}$, including relatively low-mass galaxies to examine the mass resolution problem, except in the case of Panels (d) of the two figures. 
The shaded areas show the low-mass region ($M_* < 10^{10} M_{\sun}$). 
As we mentioned above (see Section~\ref{sec:hagn_gal}), low-mass galaxies exhibit large values of PA offsets. 
This is likely a result of the fact that the current mass resolution allows only a small number of star particles and gas cells for low-mass galaxies, making it difficult to generate a realistic disk structure in which case PA alignment is difficult to achieve. 
Our mass cut of $M_* > 10^{10} M_{\sun}$ (discussed in Section~\ref{sec:hagn_gal}) is therefore justified by this argument. 
Panels (d) of the two figures hence show the Horizon-AGN galaxies above $10^{10} M_{\sun}$. 

\subsubsection{Galaxy kinematic morphology}
\label{sec:hagn_morp} 

One remarkable result is a strong correlation between star-gas misalignment and galaxy morphology, as shown in Figs.~\ref{fig:image_HAGN_phy}-(a) and \ref{fig:image_HAGN_2Dphy}-(a). 
Galaxies with lower $V/{\sigma}$ ratios tend to be more misaligned, in agreement with observations. 
In the figures, the morphology criterion ($V/{\sigma}= 1$) is expressed as a black dashed line. 
Note that extremely slow-rotating galaxies ($V/{\sigma}< 0.1$) must be treated with care. 
These galaxies are dispersion-dominated systems and their stellar rotational axes are not well defined. 
However, excluding the galaxies with $V/{\sigma}< 0.1$ does not affect our result.

Higher $V/{\sigma}$ galaxies, or LTGs, have rotation-dominated stellar component and usually have lots of gas. These properties may be linked with the misalignment fraction. 
In this subsection, we will focus on the kinematic morphology first. The effect of the gas content will be covered in the next subsection.

The dynamical settling time in relation to the ellipticity (${\epsilon} = 1-C/A$, where A and C are intrinsic major and minor axes, respectively) of galaxies may explain part of the different misalignment fractions between LTGs and ETGs. 
Fast-rotators tend to have higher ellipticities than slow-rotators \citep[e.g., the spin-ellipticity distribution.][]{2007MNRAS.379..401E}. 
\cite{2019MNRAS.483..458B} suggested that the shape of the stellar mass distribution affects the dynamical settling time (or decaying time) of the PA offset, which is the time needed for the PA offset to become 
aligned or counter-rotating. 
The rotating gas disk should be affected by gravitational torque from the stellar mass distribution and gradually become aligned with the stellar disk. 
\cite{2019MNRAS.483..458B} presented the settling time as:
\begin{equation}
    t_{\rm s} \propto \frac{R}{V_{\rm rot} (2\epsilon - \epsilon^2 ) |\cos(\phi)|},
	\label{eq:settime}
\end{equation}
where $V_{\rm rot}$ is the rotational velocity of a gas disk, R is the radius of the disk, and $\phi$ is the inclination angle between the two disks \citep[see also][]{Davis+16}.
Equation~(\ref{eq:settime}) states that the settling time increases with ellipticity and misalignment angle. 

Based on this equation, \cite{2019MNRAS.483..458B} suggested that the intrinsic ellipticity alone makes the settling-timescale for ETGs 2.7 times larger than that for LTGs. 
This estimate assumes that there is no morphological dependence on $R/V_{\rm rot}$ and that the ellipticities of ETGs and LTGs are 0.2 and 0.8, respectively.
Since the difference in the misalignment fraction between ETGs and LTGs is about 6.3 times in the SAMI sample, they concluded that the effect of ellipticity on dynamical settling time alone (i.e., Equation~(\ref{eq:settime})) is insufficient to explain the difference in misalignment fraction.
Meanwhile, Horizon-AGN presents a factor of 3.6 between ETGs and LTGs. Again, this difference is too high to be driven by the ellipticity in dynamical settling time (2.7 times) alone.

\subsubsection{The gas contents}

\label{sec:hagn_fgas} 

Misalignment is strongly affected by the gas contents of galaxies, as shown in Figs.~\ref{fig:image_HAGN_phy}-(b), -(c) and ~\ref{fig:image_HAGN_2Dphy}-(b), -(c).
Overall, the galaxies containing a smaller amount of gas more often show star-gas misalignment. 
While the gas fraction and $V/\sigma$ (morphology) are closely related, we find that they independently affect the misalignment fraction, as shown in Fig.~\ref{fig:image_HAGN_2Dphy}-(d). 
The smaller the two parameters are, the higher the misalignment fraction is.


We first focus on the impact of the amount of gas. 
To begin with, as the gas fraction increases, the momentum dissipation due to the interaction between stars and gas increases as well, resulting in a quicker decay of misalignment between them.

In addition, galaxies maintain their gas kinematic properties (e.g., direction and magnitude of spin) better when richer in gas. 
For example, the gas kinematic properties of a galaxy are influenced by gas accretion, but if the galaxy is already gas-rich, the impact of external gas accretion would naturally be small. 
Considering this effect alone, PA offset due to external gas accretion is expected to be smaller when a galaxy is richer in gas. 

While the gas mass can partially explain the trend, we also find that a good fraction of the misalignment formation is linked with gas loss (e.g., gas stripping due to the group/cluster environments). We will discuss the gas stripping in Section~\ref{sec:dis_env}, and the origin of the misalignment in the follow-up paper.


Finally, star formation in the gas disk may make the gas and stellar disks appear gradually more aligned as new stars born with the kinematic characteristics of the gas disk are added to the existing stars, affecting the mean properties of the stellar distribution.



\subsubsection{Galaxy mass}
\label{sec:hagn_masss} 

One of the important factors governing the properties of a galaxy is its mass. Therefore, the misalignment may be affected by the masses of galaxies.
Fig.~\ref{fig:image_HAGN_2Dphy}-(a) shows that more massive galaxies show higher values of misalignment fractions, which is largely due to the fact that more massive galaxies tend to be earlier in type ($V/\sigma \leq 1$). 
In order to clarify the effect of stellar mass upon the misalignment fraction, we present Fig.~\ref{fig:image_detec} here.
The green line shows the misalignment fraction of the Horizon-AGN galaxy sample. 
For comparison, we plot the SAMI misalignment fractions as a gray dotted line.
The more massive galaxies have higher misalignment fractions. 
To investigate the effects of mass and morphology, we divide the Horizon-AGN galaxies into ETGs (red) and LTGs (blue).
In the low-mass region $M_* < 10^{11}$, the green line is located near the blue line, since LTGs numerically overwhelm ETGs.
On the other hand, the green line follows the red line in the relatively high-mass region.
Therefore, the mass trend comes from a combination of massive galaxies being more likely to be ETGs \citep[e.g.,][]{2006MNRAS.373.1389C, 2010ApJ...709..644I, 2010ApJ...719.1969B}, and ETGs showing misalignment more often than LTGs.
Note that the ETG sample itself shows a mass dependence (red curve). This is probably a result of the trend that more massive early types have lower values of gas fractions and $V/\sigma$ ratios.
\cite{2019MNRAS.483..458B} also found this mass trend (gray line) and came to the same conclusion. 
They reported that massive galaxies have a slightly higher misalignment fraction than low-mass galaxies, but that the effect of morphology dominates over that of stellar mass.

\subsection{Group \& cluster environment effect}
\label{sec:env} 

Groups and clusters are gravitationally bound structures with many galaxies. 
The properties of galaxies in dense areas are affected by their environments, including factors such as the  morphology-density relation  \citep{1980ApJ...236..351D}, star formation quenching \citep{2003ApJ...584..210G}, and gas stripping \citep{1972ApJ...176....1G, 2000Sci...288.1617Q}.


Star-gas misalignment in Horizon-AGN shows a clear trend with not only the morphology of galaxies, but also with the environment.
This was already visible in Fig.~\ref{fig:image_SAMI_dist}: the cluster galaxies in Panel (d) had more than a factor-of-two-higher value of misalignment fraction compared to the whole sample (which is dominated by field galaxies).
Figs.~\ref{fig:image_HAGN_2Dgroup}-(a) and -(b) further demonstrate that group and cluster halos with greater masses or more members have higher values of misalignment fraction. 
While \cite{2019MNRAS.483..458B} found that halo mass had no effect upon misalignment fraction in the SAMI data, 
Horizon-AGN galaxies show a strong environmental trend.
Figs.~\ref{fig:image_HAGN_2Dgroup}-(c) and -(d) imply that the enhanced misalignment fraction in clusters is linked with the low gas fraction of the member galaxies, which might be related with the gas stripping process in cluster environments.
This issue will be discussed further in Section~\ref{sec:dis_env}.

\section{Discussion}
\label{sec:4}

\subsection{The environmental effect}

\label{sec:dis_env} 

A significant difference between SAMI and Horizon-AGN was found in dense environments. 
Section~\ref{sec:envdistrib} showed that Horizon-AGN has enhanced misalignment fractions in cluster environments, regardless of the galactic morphology. 
Also, the misalignment fraction is strongly linked with the halo mass or the number of its members (see Section~\ref{sec:env}). 
On the other hand, SAMI, as well as ATLAS 3D \citep{Davis+11} and MaNGA \citep{Jin+16}, show a different trend.
The cause of this discrepancy must be examined to understand the star-gas misalignment properly.



\subsubsection{Ram pressure}

The gas of cluster galaxies can be influenced by interactions with the ICM of the cluster \citep[e.g.,][]{1972ApJ...176....1G, 2000Sci...288.1617Q}, which may induce a star-gas misalignment.
Fig.~\ref{fig:image_clusterdriven} shows the properties of a typical misaligned galaxy in a cluster environment. The panel in the middle shows the infalling trajectory of this galaxy by fixing the cluster center (X mark). The virial radius ($R_{\rm 200}$) of the cluster at Stage 1 is marked with a shaded region. The PA offset (misalignment angle), shown in the color key, dramatically changes from negligible values (blue) outside of the cluster to high values (red) inside. 

The small-inset panels show the gas velocity maps of the galaxy along the trajectory. The gas rotation axis (black arrow) quickly changes once the galaxy falls inside the cluster. 
The most dramatic change in the PA offset appears at Stage 3, when the first pericenter pass of the galaxy occurs. 
The first pericenter pass is widely considered to be the place where the most dramatic ram pressure stripping occurs because both the density of the ICM and the galaxy's speed of motion attain maximal values along the trajectory.

Along with changes of the gas rotation axis, the amount of gas also systematically decreases due to ram pressure stripping, as shown in the top panel of Fig.~\ref{fig:image_clusterdriven} (green lines). 
Meanwhile, stellar components are not affected much by the cluster environment. From Stages 1 through 7, the stellar mass of the galaxy (shown in red in the top panel) hardly changes, whereas the gas mass decreases roughly by a factor of four. 
In addition, using the Yonsei Zoom-in Cluster Simulation \citep[YZiCS;][]{yzics}, \citet{2018ApJ...864...69L} found that the spin direction of the stellar rotating disk inside a cluster does not change much during its pericenter passage. We also found the same result in Horizon-AGN cluster galaxies. An independent study based on the Illustris simulation also suggested that gas stripping contributes significantly to the misalignment \citep{2019ApJ...878..143S}.
Therefore, it seems that {\em the star-gas misalignment in cluster galaxies experiencing strong gas stripping comes mainly from the ``wobbling'' gas disk} 

If the misalignment in the cluster environment is largely due to the ram pressure effect, it is expected to be correlated with halo mass and galaxy mass: a larger halo mass boosts the ram pressure effect while a larger galaxy mass counteracts it with its larger restoring force.
In Fig.~\ref{fig:image_HAGN_2Dgroup}, we see that the misalignment fraction shows a clear positive trend with halo mass. 
Using YZiCS, \cite{2018ApJ...865..156J} found that ram pressure is at work even in small group size halos, albeit at a low level. 
Consequently, the gradual change of the misalignment fraction of cluster galaxies with respect to cluster mass is physically expected.

Fig.~\ref{fig:image_detec_cluster} checks the mass trend for {\em cluster} galaxies in the same manner as Fig.~\ref{fig:image_detec}. 
A positive mass trend is visible among the cluster galaxies in the simulation when $M_* \gtrsim 10^{11}$.
We may interpret this as a result of the mass-morphology relation, as for Fig.~\ref{fig:image_detec}. 

The mass trend is reversed in the simulation galaxies at $M_* \lesssim 10^{11}$. Horizon-AGN cluster galaxies show the same mass dependence regardless of morphology (red and blue lines); hence the inverse-mass trend (green line) at $M_* \lesssim 10^{11}$ cannot be a result of the morphology mix as a function of mass.

We instead interpret this as a result of the lower restoring force of lower-mass galaxies acting against the ram pressure inside the cluster. 
These are all consistent with expectations based on the impact of ram pressure. 

The origin of the wobbling of the gas disks of cluster galaxies is unclear. 
The direct blow-away effect of ram pressure may change the gas distribution and possibly its velocity map as well, but most of the effect would be visible in the outskirts of the galaxy, where gas density is too low to be decisive on the measurement of the gas disk kinematics. 

\subsubsection{Phase-space analysis and misalignment}

The location of cluster galaxies in the phase-space diagram is known to be closely linked with the infalling history \citep[e.g.,][]{2005MNRAS.356.1327G, 2013MNRAS.431.2307O, 2017ApJ...843..128R} and the star formation activity of galaxies \citep[e.g.,][]{2014MNRAS.438.2186H, 2014ApJ...796...65M,2016MNRAS.463.3083O, 2017ApJ...843..128R, 2019ApJ...873...52O}. 

We plot a phase-space diagram in Fig.~\ref{fig:image_cluster_multi}-(a) using the clustocentric velocities and distances of galaxies. The distance axis is normalized by the virial radius of cluster ($R_{\rm 200}$), and the 3D clustocentric velocity axis is normalized by the spatial velocity dispersion of the cluster ($\sigma_{\rm 3D}$).
We define the center of the dark matter halo as the cluster center.
The cluster galaxies are color-coded based on their PA offset, and their size scales with this offset.
Overall, the values of PA offset and misalignment fraction are found to be higher in the cluster's central region than in the outskirts. 
Fig.~\ref{fig:image_cluster_multi}-(b) shows a 2D histogram of the misalignment fraction based on Panel (a) by stacking the recent 20 snapshots ($z=0 - 0.5$). 
The regions inside the virial radius ($R_{\rm 200}$) have higher misalignment fractions than those in the outskirts. 
Also, the ``backsplash region'' (BS) is found to have a higher misalignment fraction than the ``first infalling region'' (FI). Readers are referred to \citet{2017ApJ...843..128R} for more details on the phase-space analysis.

A similar trend is found in \cite{2018MNRAS.476.4753J}.
They displayed jellyfish galaxies in the phase-space diagram using the data from \cite{2016AJ....151...78P} and the GASP survey \citep{2017MNRAS.468.3883P}. 
Jellyfish galaxies often show star formation in the stripped gas, which can only be observed when fairly dense cold gas is stripped out. 
\cite{2018MNRAS.476.4753J} reported that many jellyfish galaxies are found on the left side of the phase-space diagram, which agrees well with Fig.~\ref{fig:image_cluster_multi}-(b). Also, \cite{2019MNRAS.483..458B} have found an example of the misaligned galaxies in the cluster environment with a ram pressure stripped feature. They found that the little amount of gas left in the central part of the galaxy showed a clear velocity gradient aligned with the expected gas velocity gradient due to the ram pressure stripping (see their Figure 12).


\subsection{Discrepancy between observations and simulations}
\label{sec:diff_obs_sim}
As mentioned at the beginning of Section~\ref{sec:dis_env}, an outstanding issue is that there is a serious discrepancy between observations and simulations regarding the misalignment fraction among cluster galaxies. 
While observations show no clear differences between cluster galaxies and field galaxies, Horizon-AGN shows elevated by a factor of 3 among cluster galaxies. 
We find it difficult to reconcile this tension. 

We used a different halo mass range from that of SAMI, because Horizon-AGN does not have as many massive clusters as in the observation.
Given that the misalignment fraction shows a clear positive trend with halo mass (Fig.~\ref{fig:image_HAGN_2Dgroup}), we expect the misalignment fraction to be increased, if we had as many massive clusters as in the observation.
Therefore, there would still be large discrepancy between the simulation and SAMI. 

We first suspected the grid-locking effect, which is an artifact of the grid-based calculations such as {\sc{ramses}}. The grids in the calculation are set following the distribution of stars and the gas in the first place, but the following motions of stars and the gas are affected by the grids themselves. This may affect our measurement of the motions of stars and the gas. We present our analysis on this effect in Appendix \ref{sec:AppA}. In summary, the effect does have an impact in the sense that any onset of misalignment feels friction from the pre-set grids, and hence the degree of misalignment measured in our investigation is likely a lower limit, if viewed from this perspective alone. The impact of grid-locking seems to be the same regardless of environment, providing no clue as to why we see such a large discrepancy between SAMI and Horizon-AGN on cluster galaxies. Readers are referred to Appendix \ref{sec:AppA} for more detailed information.

One possibility is that the force-calculation and mass resolutions of gas cells in the simulations are too poor to model the relevant gas distribution within cluster galaxies. 
The force-calculation resolution of Horizon-AGN ($\sim 1$ kpc) is comparable to those of the other cosmological large-volume simulations of galaxy formation; e.g., Illustris \citep{Illustris} and Eagle \citep{Eagle}. 
However, it is still much larger than the vertical scale of galactic thin disks. 
With such a low resolution, we cannot resolve the detailed features of multiphase ISM and the Kelvin-Helmholtz instability, which happens at the front of galactic disk and the ICM \citep[see also ][]{2018ApJ...865..156J}.

Cluster galaxies are relatively poorer in gas content than field galaxies. When the amount of gas is so small, it is more difficult to model properly the thin gas disk as found in real galaxies. 
When such a small amount of gas is spread out in larger areas in the simulated galaxies, it might be more vulnerable to the ram pressure stripping. This might be related with the so-called ``satellite overquenching problem'' \citep{2009MNRAS.394.1131K}.

Another possibility is that the method used to measure gas properties differs between observations and simulations. 
In our simulational analysis, we derived the gas disk properties by measuring the net angular momentum of gas within an effective radius, naively counting all gas cells/particles. 
In reality, however, observers measure the gas properties 
taking into account column density, optical depth, extinction, etc. 
The definition of ``cold gas'' also matters. IFS typically determines the gas rotation axis based on the {\em ionized} gas distribution, whereas we utilized the whole cold gas (by the density-temperature criterion) distribution in the Horizon-AGN simulation (see Section~\ref{sec:hagn_gas}).
The ionized gas of a galaxy could be misaligned from the total cold gas distribution for the following argument.
Compared to the neutral gas, ionized gas must be geographically more closely associated with young stars, and young stars form in dense regions which are less affected by environmental effects such as ram pressure. 
Moreover, PA offset can be measured differently depending on where it is based, as a significant number of galaxies show different values of PA offset at different radial distances. 
When we changed the position of measurement in the simulation data from $1\,R_{\rm eff}$ to $2\,R_{\rm eff}$, however, its impact upon the result was found to be negligible (the misalignment fraction became 10\% instead of 11\%). 
It is necessary to measure gas in the simulations more realistically by first generating mock images of galaxies and following the same measurement techniques that were used by the observers. 


Neither of these possibilities fall within the scope of the current investigation, but both are interesting subjects for future research.

\section{Summary} 
\label{sec:5}

We used the Horizon-AGN simulation to investigate the {\em properties} of star-gas misaligned galaxies. 
Overall, Horizon-AGN reproduced the observed/expected misalignment features in terms of morphology ($V/{\sigma}$), cold gas fraction, and galaxy mass, but not the observed diversity found in different environments. We summarize our results and their implications here.

We have compared the misaligned galaxies from Horizon-AGN and SAMI, applying a stellar mass cut and a cold gas detection limit.
Horizon-AGN reproduced the PA offset distribution of SAMI galaxies reasonably well. 
However, Horizon-AGN did not reproduce the small peaks at 90- and 180-degrees observed by SAMI.
It is probable that the force-calculation resolution of Horizon-AGN is insufficient to resolve such small peaks or that the resolution causes small peaks dissolve to more quickly in simulations than in real galaxies. 
 
ETGs are found to show larger misalignment fractions both in SAMI and Horizon-AGN.
Horizon-AGN galaxies with lower values of $V/{\sigma}$ (kinematic morphology indicator) tend to have higher misalignment fractions and higher values of PA offsets. While the dynamical settling time depending on ellipticity can partially explain this phenomenon, it is insufficient to explain all of it.
 
We found in the Horizon-AGN galaxies that kinematic morphology and gas fraction independently affect the misalignment fraction. 
Smaller values of the two parameters ($V/{\sigma}$ and cold gas fraction) correspond to higher misalignment fractions.
Galaxies with higher gas fractions can sustain their gas kinematics more easily against external gas accretion. 
 
The misalignment fraction is also seemingly affected by stellar mass.
However, we have found that this trend largely arises from the fact that more massive galaxies tend to be earlier in type with lower in gas content. 

Recently, \citet{Duckworth+20} performed a similar work to this study using the MaNGA observations and the IllustrisTNG100 simulation and found consistent results.

One outstanding discrepancy between observations (SAMI) and simulations (Horizon-AGN) was found in dense (cluster) environments. 
Observations found no clear difference in misalignment fraction between field and cluster environments, whereas Horizon-AGN found a factor of three higher values in cluster galaxies regardless of morphology.
This enhanced misalignment fraction in Horizon-AGN also shows a strong correlation with halo mass. We found that star-gas misalignment in cluster galaxies experiencing strong ram pressure stripping comes mainly from the wobbling gas disk.
We suspect that the low force-calculation resolution of current large-volume simulations and/or the use of different gas measurement techniques contribute significantly to the discrepancy.


We will investigate the {\em origin and evolution} of misaligned galaxies, identifying different channels of misalignment formation and quantifying their levels of significance, in the follow-up paper, Paper II (Khim et al. in prep). 
We will also measure the lifetimes of star-gas misalignment for different types of galaxies.

\acknowledgments

We thank the referee for constructive comments that significantly improved the clarity of the manuscript. SKY acted as the corresponding author and acknowledges support from the Korean National Research Foundation (NRF-2017R1A2A05001116). 
DJK acknowledges support from Yonsei University through Yonsei Honors Scholarship. 
JJB acknowledges support of an Australian Research Council Future Fellowship (FT180100231).
JBH is supported by an ARC Laureate Fellowship that funds Jesse van de Sande and an ARC Federation Fellowship that funded the SAMI prototype.
MSO acknowledges the funding support from the Australian Research Council through a Future Fellowship (FT140100255). 
JvdS is funded under Bland-Hawthorn's ARC Laureate Fellowship (FL140100278). 
Parts of this research were conducted by the Australian Research Council Centre of Excellence for All Sky Astrophysics in 3 Dimensions (ASTRO 3D), through project number CE170100013.
This work relied on the HPC resources of the Horizon Cluster hosted by Institut d’Astrophysique de Paris. We warmly thank S. Rouberol for running the cluster on which the simulation was post-processed.  
This work is partially supported by the Spin(e) grant ANR-13-BS05-0005 of the French Agence Nationale de la Recherche.
The SAMI Galaxy Survey is based on observations made at the Anglo-Australian Telescope. The Sydney-AAO Multi-object Integral field spectrograph (SAMI) was developed jointly by the University of Sydney and the Australian Astronomical Observatory. The SAMI input catalogue is based on data taken from the Sloan Digital Sky Survey, the GAMA Survey and the VST ATLAS Survey. The SAMI Galaxy Survey is supported by the Australian Research Council Centre of Excellence for All Sky Astrophysics in 3 Dimensions (ASTRO 3D), through project number CE170100013, the Australian Research Council Centre of Excellence for All-sky Astrophysics (CAASTRO), through project number CE110001020, and other participating institutions. The SAMI Galaxy Survey website is \url{http://sami-survey.org/}


\bibliography{ref}{}

\begin{thebibliography}{82}
\expandafter\ifx\csname natexlab\endcsname\relax\def\natexlab#1{#1}\fi

\bibitem[{{Algorry} {et~al.}(2014){Algorry}, {Navarro}, {Abadi}, {Sales},
  {Steinmetz}, \& {Piontek}}]{2014MNRAS.437.3596A}
{Algorry}, D.~G., {Navarro}, J.~F., {Abadi}, M.~G., {et~al.} 2014,
  \href{http://dx.doi.org/10.1093/mnras/stt2154}{\color{magenta}\mnras},
  \href{https://ui.adsabs.harvard.edu/abs/2014MNRAS.437.3596A}{437, 3596}

\bibitem[{{Aubert} {et~al.}(2004){Aubert}, {Pichon}, \&
  {Colombi}}]{2004MNRAS.352..376A}
{Aubert}, D., {Pichon}, C., \& {Colombi}, S. 2004,
  \href{http://dx.doi.org/10.1111/j.1365-2966.2004.07883.x}{\color{magenta}\mnras},
  \href{http://adsabs.harvard.edu/abs/2004MNRAS.352..376A}{352, 376}

\bibitem[{{Aumer} \& {White}(2013)}]{2013MNRAS.428.1055A}
{Aumer}, M. \& {White}, S. D.~M. 2013,
  \href{http://dx.doi.org/10.1093/mnras/sts083}{\color{magenta}\mnras},
  \href{https://ui.adsabs.harvard.edu/abs/2013MNRAS.428.1055A}{428, 1055}

\bibitem[{{Balcells} \& {Quinn}(1990)}]{1990ApJ...361..381B}
{Balcells}, M. \& {Quinn}, P.~J. 1990,
  \href{http://dx.doi.org/10.1086/169204}{\color{magenta}\apj},
  \href{https://ui.adsabs.harvard.edu/abs/1990ApJ...361..381B}{361, 381}

\bibitem[{{Barnes} \& {Hernquist}(1996)}]{1996ApJ...471..115B}
{Barnes}, J.~E. \& {Hernquist}, L. 1996,
  \href{http://dx.doi.org/10.1086/177957}{\color{magenta}\apj},
  \href{https://ui.adsabs.harvard.edu/abs/1996ApJ...471..115B}{471, 115}

\bibitem[{{Barrera-Ballesteros} {et~al.}(2014){Barrera-Ballesteros},
  {Falc{\'o}n-Barroso}, {Garc{\'{\i}}a-Lorenzo}, {van de Ven}, {Aguerri},
  {Mendez-Abreu}, {Spekkens}, {Lyubenova}, {S{\'a}nchez}, {Husemann}, {Mast},
  {Garc{\'{\i}}a-Benito}, {Iglesias-Paramo}, {Del Olmo}, {M{\'a}rquez},
  {Masegosa}, {Kehrig}, {Marino}, {Verdes-Montenegro}, {Ziegler}, {McIntosh},
  {Bland-Hawthorn}, {Walcher}, \& {Califa Collaboration}}]{2014A&A...568A..70B}
{Barrera-Ballesteros}, J.~K., {Falc{\'o}n-Barroso}, J.,
  {Garc{\'{\i}}a-Lorenzo}, B., {et~al.} 2014,
  \href{http://dx.doi.org/10.1051/0004-6361/201423488}{\color{magenta}\aap},
  \href{http://adsabs.harvard.edu/abs/2014A%26A...568A..70B}{568, A70}

\bibitem[{{Barrera-Ballesteros} {et~al.}(2015){Barrera-Ballesteros},
  {Garc{\'{\i}}a-Lorenzo}, {Falc{\'o}n-Barroso}, {van de Ven}, {Lyubenova},
  {Wild}, {M{\'e}ndez-Abreu}, {S{\'a}nchez}, {Marquez}, {Masegosa},
  {Monreal-Ibero}, {Ziegler}, {del Olmo}, {Verdes-Montenegro},
  {Garc{\'{\i}}a-Benito}, {Husemann}, {Mast}, {Kehrig}, {Iglesias-Paramo},
  {Marino}, {Aguerri}, {Walcher}, {V{\'{\i}}lchez}, {Bomans},
  {Cortijo-Ferrero}, {Gonz{\'a}lez Delgado}, {Bland-Hawthorn}, {McIntosh}, \&
  {Bekerait{\.e}}}]{2015A&A...582A..21B}
{Barrera-Ballesteros}, J.~K., {Garc{\'{\i}}a-Lorenzo}, B.,
  {Falc{\'o}n-Barroso}, J., {et~al.} 2015,
  \href{http://dx.doi.org/10.1051/0004-6361/201424935}{\color{magenta}\aap},
  \href{http://adsabs.harvard.edu/abs/2015A%26A...582A..21B}{582, A21}

\bibitem[{{Bekki}(1998)}]{1998ApJ...499..635B}
{Bekki}, K. 1998, \href{http://dx.doi.org/10.1086/305680}{\color{magenta}\apj},
  \href{https://ui.adsabs.harvard.edu/abs/1998ApJ...499..635B}{499, 635}

\bibitem[{{Bertola} {et~al.}(1992){Bertola}, {Buson}, \&
  {Zeilinger}}]{1992ApJ...401L..79B}
{Bertola}, F., {Buson}, L.~M., \& {Zeilinger}, W.~W. 1992,
  \href{http://dx.doi.org/10.1086/186675}{\color{magenta}\apj},
  \href{https://ui.adsabs.harvard.edu/abs/1992ApJ...401L..79B}{401, L79}

\bibitem[{{Bournaud} \& {Combes}(2003)}]{2003A&A...401..817B}
{Bournaud}, F. \& {Combes}, F. 2003,
  \href{http://dx.doi.org/10.1051/0004-6361:20030150}{\color{magenta}\aap},
  \href{https://ui.adsabs.harvard.edu/abs/2003A&A...401..817B}{401, 817}

\bibitem[{{Brook} {et~al.}(2008){Brook}, {Governato}, {Quinn}, {Wadsley},
  {Brooks}, {Willman}, {Stilp}, \& {Jonsson}}]{2008ApJ...689..678B}
{Brook}, C.~B., {Governato}, F., {Quinn}, T., {et~al.} 2008,
  \href{http://dx.doi.org/10.1086/591489}{\color{magenta}\apj},
  \href{https://ui.adsabs.harvard.edu/abs/2008ApJ...689..678B}{689, 678}

\bibitem[{{Bryant} {et~al.}(2019){Bryant}, {Croom}, {van de Sande}, {Scott},
  {Fogarty}, {Bland-Hawthorn}, {Bloom}, {Taylor}, {Brough}, {Robotham},
  {Cortese}, {Couch}, {Owers}, {Medling}, {Federrath}, {Bekki}, {Richards},
  {Lawrence}, \& {Konstantopoulos}}]{2019MNRAS.483..458B}
{Bryant}, J.~J., {Croom}, S.~M., {van de Sande}, J., {et~al.} 2019,
  \href{http://dx.doi.org/10.1093/mnras/sty3122}{\color{magenta}\mnras},
  \href{http://adsabs.harvard.edu/abs/2019MNRAS.483..458B}{483, 458}

\bibitem[{{Bryant} {et~al.}(2015){Bryant}, {Owers}, {Robotham}, {Croom},
  {Driver}, {Drinkwater}, {Lorente}, {Cortese}, {Scott}, {Colless}, {Schaefer},
  {Taylor}, {Konstantopoulos}, {Allen}, {Baldry}, {Barnes}, {Bauer},
  {Bland-Hawthorn}, {Bloom}, {Brooks}, {Brough}, {Cecil}, {Couch}, {Croton},
  {Davies}, {Ellis}, {Fogarty}, {Foster}, {Glazebrook}, {Goodwin}, {Green},
  {Gunawardhana}, {Hampton}, {Ho}, {Hopkins}, {Kewley}, {Lawrence},
  {Leon-Saval}, {Leslie}, {McElroy}, {Lewis}, {Liske}, {L{\'o}pez-S{\'a}nchez},
  {Mahajan}, {Medling}, {Metcalfe}, {Meyer}, {Mould}, {Obreschkow}, {O'Toole},
  {Pracy}, {Richards}, {Shanks}, {Sharp}, {Sweet}, {Thomas}, {Tonini}, \&
  {Walcher}}]{2015MNRAS.447.2857B}
{Bryant}, J.~J., {Owers}, M.~S., {Robotham}, A.~S.~G., {et~al.} 2015,
  \href{http://dx.doi.org/10.1093/mnras/stu2635}{\color{magenta}\mnras},
  \href{http://adsabs.harvard.edu/abs/2015MNRAS.447.2857B}{447, 2857}

\bibitem[{{Bundy} {et~al.}(2010){Bundy}, {Scarlata}, {Carollo}, {Ellis},
  {Drory}, {Hopkins}, {Salvato}, {Leauthaud}, {Koekemoer}, {Murray}, {Ilbert},
  {Oesch}, {Ma}, {Capak}, {Pozzetti}, \& {Scoville}}]{2010ApJ...719.1969B}
{Bundy}, K., {Scarlata}, C., {Carollo}, C.~M., {et~al.} 2010,
  \href{http://dx.doi.org/10.1088/0004-637X/719/2/1969}{\color{magenta}\apj},
  \href{http://adsabs.harvard.edu/abs/2010ApJ...719.1969B}{719, 1969}

\bibitem[{{Bureau} \& {Chung}(2006)}]{2006MNRAS.366..182B}
{Bureau}, M. \& {Chung}, A. 2006,
  \href{http://dx.doi.org/10.1111/j.1365-2966.2005.09840.x}{\color{magenta}\mnras},
  \href{https://ui.adsabs.harvard.edu/abs/2006MNRAS.366..182B}{366, 182}

\bibitem[{{Cappellari} \& {Emsellem}(2004)}]{2004PASP..116..138C}
{Cappellari}, M. \& {Emsellem}, E. 2004,
  \href{http://dx.doi.org/10.1086/381875}{\color{magenta}\pasp},
  \href{http://adsabs.harvard.edu/abs/2004PASP..116..138C}{116, 138}

\bibitem[{{Chisari} {et~al.}(2015){Chisari}, {Codis}, {Laigle}, {Dubois},
  {Pichon}, {Devriendt}, {Slyz}, {Miller}, {Gavazzi}, \&
  {Benabed}}]{2015MNRAS.454.2736C}
{Chisari}, N., {Codis}, S., {Laigle}, C., {et~al.} 2015,
  \href{http://dx.doi.org/10.1093/mnras/stv2154}{\color{magenta}\mnras},
  \href{https://ui.adsabs.harvard.edu/abs/2015MNRAS.454.2736C}{454, 2736}

\bibitem[{{Choi} \& {Yi}(2017)}]{yzics}
{Choi}, H. \& {Yi}, S.~K. 2017,
  \href{http://dx.doi.org/10.3847/1538-4357/aa5e4b}{\color{magenta}\apj},
  \href{http://adsabs.harvard.edu/abs/2017ApJ...837...68C}{837, 68}

\bibitem[{{Chung} {et~al.}(2006){Chung}, {Koribalski}, {Bureau}, \& {van
  Gorkom}}]{2006MNRAS.370.1565C}
{Chung}, A., {Koribalski}, B., {Bureau}, M., \& {van Gorkom}, J.~H. 2006,
  \href{http://dx.doi.org/10.1111/j.1365-2966.2006.10579.x}{\color{magenta}\mnras},
  \href{https://ui.adsabs.harvard.edu/abs/2006MNRAS.370.1565C}{370, 1565}

\bibitem[{{Coccato} {et~al.}(2015){Coccato}, {Fabricius}, {Morelli}, {Corsini},
  {Pizzella}, {Erwin}, {Dalla Bont{\`a}}, {Saglia}, {Bender}, \&
  {Williams}}]{2015A&A...581A..65C}
{Coccato}, L., {Fabricius}, M., {Morelli}, L., {et~al.} 2015,
  \href{http://dx.doi.org/10.1051/0004-6361/201526560}{\color{magenta}\aap},
  \href{https://ui.adsabs.harvard.edu/abs/2015A&A...581A..65C}{581, A65}

\bibitem[{{Coccato} {et~al.}(2011){Coccato}, {Morelli}, {Corsini}, {Buson},
  {Pizzella}, {Vergani}, \& {Bertola}}]{2011MNRAS.412L.113C}
{Coccato}, L., {Morelli}, L., {Corsini}, E.~M., {et~al.} 2011,
  \href{http://dx.doi.org/10.1111/j.1745-3933.2011.01016.x}{\color{magenta}\mnras},
  \href{https://ui.adsabs.harvard.edu/abs/2011MNRAS.412L.113C}{412, L113}

\bibitem[{{Conselice}(2006)}]{2006MNRAS.373.1389C}
{Conselice}, C.~J. 2006,
  \href{http://dx.doi.org/10.1111/j.1365-2966.2006.11114.x}{\color{magenta}\mnras},
  \href{http://adsabs.harvard.edu/abs/2006MNRAS.373.1389C}{373, 1389}

\bibitem[{{Cortese} {et~al.}(2016){Cortese}, {Fogarty}, {Bekki}, {van de
  Sande}, {Couch}, {Catinella}, {Colless}, {Obreschkow}, {Taranu}, {Tescari},
  {Barat}, {Bland-Hawthorn}, {Bloom}, {Bryant}, {Cluver}, {Croom},
  {Drinkwater}, {d'Eugenio}, {Konstantopoulos}, {Lopez-Sanchez}, {Mahajan},
  {Scott}, {Tonini}, {Wong}, {Allen}, {Brough}, {Goodwin}, {Green}, {Ho},
  {Kelvin}, {Lawrence}, {Lorente}, {Medling}, {Owers}, {Richards}, {Sharp}, \&
  {Sweet}}]{2016MNRAS.463..170C}
{Cortese}, L., {Fogarty}, L.~M.~R., {Bekki}, K., {et~al.} 2016,
  \href{http://dx.doi.org/10.1093/mnras/stw1891}{\color{magenta}\mnras},
  \href{http://adsabs.harvard.edu/abs/2016MNRAS.463..170C}{463, 170}

\bibitem[{{Crocker} {et~al.}(2009){Crocker}, {Jeong}, {Komugi}, {Combes},
  {Bureau}, {Young}, \& {Yi}}]{2009MNRAS.393.1255C}
{Crocker}, A.~F., {Jeong}, H., {Komugi}, S., {et~al.} 2009,
  \href{http://dx.doi.org/10.1111/j.1365-2966.2008.14295.x}{\color{magenta}\mnras},
  \href{https://ui.adsabs.harvard.edu/abs/2009MNRAS.393.1255C}{393, 1255}

\bibitem[{{Croom} {et~al.}(2012){Croom}, {Lawrence}, {Bland-Hawthorn},
  {Bryant}, {Fogarty}, {Richards}, {Goodwin}, {Farrell}, {Miziarski}, {Heald},
  {Jones}, {Lee}, {Colless}, {Brough}, {Hopkins}, {Bauer}, {Birchall}, {Ellis},
  {Horton}, {Leon-Saval}, {Lewis}, {L{\'o}pez-S{\'a}nchez}, {Min}, {Trinh}, \&
  {Trowland}}]{2012MNRAS.421..872C}
{Croom}, S.~M., {Lawrence}, J.~S., {Bland-Hawthorn}, J., {et~al.} 2012,
  \href{http://dx.doi.org/10.1111/j.1365-2966.2011.20365.x}{\color{magenta}\mnras},
  \href{http://adsabs.harvard.edu/abs/2012MNRAS.421..872C}{421, 872}

\bibitem[{{Davis} {et~al.}(2011){Davis}, {Alatalo}, {Sarzi}, {Bureau}, {Young},
  {Blitz}, {Serra}, {Crocker}, {Krajnovi{\'c}}, {McDermid}, {Bois}, {Bournaud},
  {Cappellari}, {Davies}, {Duc}, {de Zeeuw}, {Emsellem}, {Khochfar},
  {Kuntschner}, {Lablanche}, {Morganti}, {Naab}, {Oosterloo}, {Scott}, \&
  {Weijmans}}]{Davis+11}
{Davis}, T.~A., {Alatalo}, K., {Sarzi}, M., {et~al.} 2011,
  \href{http://dx.doi.org/10.1111/j.1365-2966.2011.19355.x}{\color{magenta}\mnras},
  \href{http://adsabs.harvard.edu/abs/2011MNRAS.417..882D}{417, 882}

\bibitem[{{Davis} \& {Bureau}(2016)}]{Davis+16}
{Davis}, T.~A. \& {Bureau}, M. 2016,
  \href{http://dx.doi.org/10.1093/mnras/stv2998}{\color{magenta}\mnras},
  \href{http://adsabs.harvard.edu/abs/2016MNRAS.457..272D}{457, 272}

\bibitem[{{De Rijcke} {et~al.}(2004){De Rijcke}, {Dejonghe}, {Zeilinger}, \&
  {Hau}}]{2004A&A...426...53D}
{De Rijcke}, S., {Dejonghe}, H., {Zeilinger}, W.~W., \& {Hau}, G.~K.~T. 2004,
  \href{http://dx.doi.org/10.1051/0004-6361:20041205}{\color{magenta}\aap},
  \href{https://ui.adsabs.harvard.edu/abs/2004A&A...426...53D}{426, 53}

\bibitem[{{Dressler}(1980)}]{1980ApJ...236..351D}
{Dressler}, A. 1980,
  \href{http://dx.doi.org/10.1086/157753}{\color{magenta}\apj},
  \href{http://adsabs.harvard.edu/abs/1980ApJ...236..351D}{236, 351}

\bibitem[{{Dubois} {et~al.}(2016){Dubois}, {Peirani}, {Pichon}, {Devriendt},
  {Gavazzi}, {Welker}, \& {Volonteri}}]{2016MNRAS.463.3948D}
{Dubois}, Y., {Peirani}, S., {Pichon}, C., {et~al.} 2016,
  \href{http://dx.doi.org/10.1093/mnras/stw2265}{\color{magenta}\mnras},
  \href{http://adsabs.harvard.edu/abs/2016MNRAS.463.3948D}{463, 3948}

\bibitem[{{Dubois} {et~al.}(2014){Dubois}, {Pichon}, {Welker}, {Le Borgne},
  {Devriendt}, {Laigle}, {Codis}, {Pogosyan}, {Arnouts}, {Benabed}, {Bertin},
  {Blaizot}, {Bouchet}, {Cardoso}, {Colombi}, {de Lapparent}, {Desjacques},
  {Gavazzi}, {Kassin}, {Kimm}, {McCracken}, {Milliard}, {Peirani}, {Prunet},
  {Rouberol}, {Silk}, {Slyz}, {Sousbie}, {Teyssier}, {Tresse}, {Treyer},
  {Vibert}, \& {Volonteri}}]{2014MNRAS.444.1453D}
{Dubois}, Y., {Pichon}, C., {Welker}, C., {et~al.} 2014,
  \href{http://dx.doi.org/10.1093/mnras/stu1227}{\color{magenta}\mnras},
  \href{http://adsabs.harvard.edu/abs/2014MNRAS.444.1453D}{444, 1453}

\bibitem[{{Duckworth} {et~al.}(2020){Duckworth}, {Tojeiro}, \&
  {Kraljic}}]{Duckworth+20}
{Duckworth}, C., {Tojeiro}, R., \& {Kraljic}, K. 2020,
  \href{http://dx.doi.org/10.1093/mnras/stz3575}{\color{magenta}\mnras},
  \href{https://ui.adsabs.harvard.edu/abs/2020MNRAS.492.1869D}{492, 1869}

\bibitem[{{Emsellem} {et~al.}(2007){Emsellem}, {Cappellari}, {Krajnovi{\'c}},
  {van de Ven}, {Bacon}, {Bureau}, {Davies}, {de Zeeuw}, {Falc{\'o}n-Barroso},
  {Kuntschner}, {McDermid}, {Peletier}, \& {Sarzi}}]{2007MNRAS.379..401E}
{Emsellem}, E., {Cappellari}, M., {Krajnovi{\'c}}, D., {et~al.} 2007,
  \href{http://dx.doi.org/10.1111/j.1365-2966.2007.11752.x}{\color{magenta}\mnras},
  \href{http://adsabs.harvard.edu/abs/2007MNRAS.379..401E}{379, 401}

\bibitem[{{Gill} {et~al.}(2005){Gill}, {Knebe}, \&
  {Gibson}}]{2005MNRAS.356.1327G}
{Gill}, S. P.~D., {Knebe}, A., \& {Gibson}, B.~K. 2005,
  \href{http://dx.doi.org/10.1111/j.1365-2966.2004.08562.x}{\color{magenta}\mnras},
  \href{https://ui.adsabs.harvard.edu/abs/2005MNRAS.356.1327G}{356, 1327}

\bibitem[{{G{\'o}mez} {et~al.}(2003){G{\'o}mez}, {Nichol}, {Miller}, {Balogh},
  {Goto}, {Zabludoff}, {Romer}, {Bernardi}, {Sheth}, {Hopkins}, {Castander},
  {Connolly}, {Schneider}, {Brinkmann}, {Lamb}, {SubbaRao}, \&
  {York}}]{2003ApJ...584..210G}
{G{\'o}mez}, P.~L., {Nichol}, R.~C., {Miller}, C.~J., {et~al.} 2003,
  \href{http://dx.doi.org/10.1086/345593}{\color{magenta}\apj},
  \href{https://ui.adsabs.harvard.edu/abs/2003ApJ...584..210G}{584, 210}

\bibitem[{{Gunn} \& {Gott}(1972)}]{1972ApJ...176....1G}
{Gunn}, J.~E. \& {Gott}, III, J.~R. 1972,
  \href{http://dx.doi.org/10.1086/151605}{\color{magenta}\apj},
  \href{http://adsabs.harvard.edu/abs/1972ApJ...176....1G}{176, 1}

\bibitem[{{Hern{\'a}ndez-Fern{\'a}ndez}
  {et~al.}(2014){Hern{\'a}ndez-Fern{\'a}ndez}, {Haines}, {Diaferio},
  {Iglesias-P{\'a}ramo}, {Mendes de Oliveira}, \&
  {Vilchez}}]{2014MNRAS.438.2186H}
{Hern{\'a}ndez-Fern{\'a}ndez}, J.~D., {Haines}, C.~P., {Diaferio}, A., {et~al.}
  2014, \href{http://dx.doi.org/10.1093/mnras/stt2354}{\color{magenta}\mnras},
  \href{https://ui.adsabs.harvard.edu/abs/2014MNRAS.438.2186H}{438, 2186}

\bibitem[{{Hernquist} \& {Barnes}(1991)}]{1991Natur.354..210H}
{Hernquist}, L. \& {Barnes}, J.~E. 1991,
  \href{http://dx.doi.org/10.1038/354210a0}{\color{magenta}\nat},
  \href{https://ui.adsabs.harvard.edu/abs/1991Natur.354..210H}{354, 210}

\bibitem[{{Ilbert} {et~al.}(2010){Ilbert}, {Salvato}, {Le Floc'h}, {Aussel},
  {Capak}, {McCracken}, {Mobasher}, {Kartaltepe}, {Scoville}, {Sanders},
  {Arnouts}, {Bundy}, {Cassata}, {Kneib}, {Koekemoer}, {Le F{\`e}vre}, {Lilly},
  {Surace}, {Taniguchi}, {Tasca}, {Thompson}, {Tresse}, {Zamojski}, {Zamorani},
  \& {Zucca}}]{2010ApJ...709..644I}
{Ilbert}, O., {Salvato}, M., {Le Floc'h}, E., {et~al.} 2010,
  \href{http://dx.doi.org/10.1088/0004-637X/709/2/644}{\color{magenta}\apj},
  \href{http://adsabs.harvard.edu/abs/2010ApJ...709..644I}{709, 644}

\bibitem[{{Jaff{\'e}} {et~al.}(2018){Jaff{\'e}}, {Poggianti}, {Moretti},
  {Gullieuszik}, {Smith}, {Vulcani}, {Fasano}, {Fritz}, {Tonnesen}, \&
  {Bettoni}}]{2018MNRAS.476.4753J}
{Jaff{\'e}}, Y.~L., {Poggianti}, B.~M., {Moretti}, A., {et~al.} 2018,
  \href{http://dx.doi.org/10.1093/mnras/sty500}{\color{magenta}\mnras},
  \href{https://ui.adsabs.harvard.edu/abs/2018MNRAS.476.4753J}{476, 4753}

\bibitem[{{Jin} {et~al.}(2016){Jin}, {Chen}, {Shi}, {Tremonti}, {Bershady},
  {Merrifield}, {Emsellem}, {Fu}, {Wake}, {Bundy}, {Lin}, {Argudo-Fernandez},
  {Huang}, {Stark}, {Storchi-Bergmann}, {Bizyaev}, {Brownstein}, {Chisholm},
  {Guo}, {Hao}, {Hu}, {Li}, {Li}, {Masters}, {Malanushenko}, {Pan}, {Riffel},
  {Roman-Lopes}, {Simmons}, {Thomas}, {Wang}, {Westfall}, \& {Yan}}]{Jin+16}
{Jin}, Y., {Chen}, Y., {Shi}, Y., {et~al.} 2016,
  \href{http://dx.doi.org/10.1093/mnras/stw2055}{\color{magenta}\mnras},
  \href{http://adsabs.harvard.edu/abs/2016MNRAS.463..913J}{463, 913}

\bibitem[{{Jung} {et~al.}(2018){Jung}, {Choi}, {Wong}, {Kimm}, {Chung}, \&
  {Yi}}]{2018ApJ...865..156J}
{Jung}, S.~L., {Choi}, H., {Wong}, O.~I., {et~al.} 2018,
  \href{http://dx.doi.org/10.3847/1538-4357/aadda2}{\color{magenta}\apj},
  \href{https://ui.adsabs.harvard.edu/abs/2018ApJ...865..156J}{865, 156}

\bibitem[{{Kannappan} \& {Fabricant}(2001)}]{2001AJ....121..140K}
{Kannappan}, S.~J. \& {Fabricant}, D.~G. 2001,
  \href{http://dx.doi.org/10.1086/318027}{\color{magenta}\aj},
  \href{https://ui.adsabs.harvard.edu/abs/2001AJ....121..140K}{121, 140}

\bibitem[{{Katkov} {et~al.}(2016){Katkov}, {Sil'chenko}, {Chilingarian},
  {Uklein}, \& {Egorov}}]{2016MNRAS.461.2068K}
{Katkov}, I.~Y., {Sil'chenko}, O.~K., {Chilingarian}, I.~V., {Uklein}, R.~I.,
  \& {Egorov}, O.~V. 2016,
  \href{http://dx.doi.org/10.1093/mnras/stw1452}{\color{magenta}\mnras},
  \href{https://ui.adsabs.harvard.edu/abs/2016MNRAS.461.2068K}{461, 2068}

\bibitem[{{Kimm} {et~al.}(2009){Kimm}, {Somerville}, {Yi}, {van den Bosch},
  {Salim}, {Fontanot}, {Monaco}, {Mo}, {Pasquali}, {Rich}, \&
  {Yang}}]{2009MNRAS.394.1131K}
{Kimm}, T., {Somerville}, R.~S., {Yi}, S.~K., {et~al.} 2009,
  \href{http://dx.doi.org/10.1111/j.1365-2966.2009.14414.x}{\color{magenta}\mnras},
  \href{https://ui.adsabs.harvard.edu/abs/2009MNRAS.394.1131K}{394, 1131}

\bibitem[{{Komatsu} {et~al.}(2011){Komatsu}, {Smith}, {Dunkley}, {Bennett},
  {Gold}, {Hinshaw}, {Jarosik}, {Larson}, {Nolta}, {Page}, {Spergel},
  {Halpern}, {Hill}, {Kogut}, {Limon}, {Meyer}, {Odegard}, {Tucker}, {Weiland},
  {Wollack}, \& {Wright}}]{2011ApJS..192...18K}
{Komatsu}, E., {Smith}, K.~M., {Dunkley}, J., {et~al.} 2011,
  \href{http://dx.doi.org/10.1088/0067-0049/192/2/18}{\color{magenta}\apjs},
  \href{http://adsabs.harvard.edu/abs/2011ApJS..192...18K}{192, 18}

\bibitem[{{Krajnovi{\'c}} {et~al.}(2015){Krajnovi{\'c}}, {Weilbacher},
  {Urrutia}, {Emsellem}, {Carollo}, {Shirazi}, {Bacon}, {Contini}, {Epinat}, \&
  {Kamann}}]{2015MNRAS.452....2K}
{Krajnovi{\'c}}, D., {Weilbacher}, P.~M., {Urrutia}, T., {et~al.} 2015,
  \href{http://dx.doi.org/10.1093/mnras/stv958}{\color{magenta}\mnras},
  \href{https://ui.adsabs.harvard.edu/abs/2015MNRAS.452....2K}{452, 2}

\bibitem[{{Kraljic} {et~al.}(2020){Kraljic}, {Dav{\'e}}, \&
  {Pichon}}]{2020MNRAS.tmp..237K}
{Kraljic}, K., {Dav{\'e}}, R., \& {Pichon}, C. 2020,
  \href{https://ui.adsabs.harvard.edu/abs/2020MNRAS.tmp..237K}{\href{http://dx.doi.org/10.1093/mnras/staa250}{\color{magenta}\mnras},
  237}

\bibitem[{{Kuijken} {et~al.}(1996){Kuijken}, {Fisher}, \&
  {Merrifield}}]{1996MNRAS.283..543K}
{Kuijken}, K., {Fisher}, D., \& {Merrifield}, M.~R. 1996,
  \href{http://dx.doi.org/10.1093/mnras/283.2.543}{\color{magenta}\mnras},
  \href{https://ui.adsabs.harvard.edu/abs/1996MNRAS.283..543K}{283, 543}

\bibitem[{{Lagos} {et~al.}(2015){Lagos}, {Padilla}, {Davis}, {Lacey}, {Baugh},
  {Gonzalez-Perez}, {Zwaan}, \& {Contreras}}]{2015MNRAS.448.1271L}
{Lagos}, C.~d.~P., {Padilla}, N.~D., {Davis}, T.~A., {et~al.} 2015,
  \href{http://dx.doi.org/10.1093/mnras/stu2763}{\color{magenta}\mnras},
  \href{http://adsabs.harvard.edu/abs/2015MNRAS.448.1271L}{448, 1271}

\bibitem[{{Lee} {et~al.}(2018){Lee}, {Kim}, {Jeong}, {Smith}, {Choi}, {Hwang},
  {Joo}, {Kim}, {Lee}, \& {Yi}}]{2018ApJ...864...69L}
{Lee}, J., {Kim}, S., {Jeong}, H., {et~al.} 2018,
  \href{http://dx.doi.org/10.3847/1538-4357/aad54e}{\color{magenta}\apj},
  \href{https://ui.adsabs.harvard.edu/abs/2018ApJ...864...69L}{864, 69}

\bibitem[{{Muzzin} {et~al.}(2014){Muzzin}, {van der Burg}, {McGee}, {Balogh},
  {Franx}, {Hoekstra}, {Hudson}, {Noble}, {Taranu}, \&
  {Webb}}]{2014ApJ...796...65M}
{Muzzin}, A., {van der Burg}, R.~F.~J., {McGee}, S.~L., {et~al.} 2014,
  \href{http://dx.doi.org/10.1088/0004-637X/796/1/65}{\color{magenta}\apj},
  \href{https://ui.adsabs.harvard.edu/abs/2014ApJ...796...65M}{796, 65}

\bibitem[{{Oman} \& {Hudson}(2016)}]{2016MNRAS.463.3083O}
{Oman}, K.~A. \& {Hudson}, M.~J. 2016,
  \href{http://dx.doi.org/10.1093/mnras/stw2195}{\color{magenta}\mnras},
  \href{https://ui.adsabs.harvard.edu/abs/2016MNRAS.463.3083O}{463, 3083}

\bibitem[{{Oman} {et~al.}(2013){Oman}, {Hudson}, \&
  {Behroozi}}]{2013MNRAS.431.2307O}
{Oman}, K.~A., {Hudson}, M.~J., \& {Behroozi}, P.~S. 2013,
  \href{http://dx.doi.org/10.1093/mnras/stt328}{\color{magenta}\mnras},
  \href{https://ui.adsabs.harvard.edu/abs/2013MNRAS.431.2307O}{431, 2307}

\bibitem[{{Osman} \& {Bekki}(2017)}]{2017MNRAS.471L..87O}
{Osman}, O. \& {Bekki}, K. 2017,
  \href{http://dx.doi.org/10.1093/mnrasl/slx104}{\color{magenta}\mnras},
  \href{https://ui.adsabs.harvard.edu/abs/2017MNRAS.471L..87O}{471, L87}

\bibitem[{{Owers} {et~al.}(2017){Owers}, {Allen}, {Baldry}, {Bryant}, {Cecil},
  {Cortese}, {Croom}, {Driver}, {Fogarty}, {Green}, {Helmich}, {de Jong},
  {Kuijken}, {Mahajan}, {McFarland}, {Pracy}, {Robotham}, {Sikkema}, {Sweet},
  {Taylor}, {Verdoes Kleijn}, {Bauer}, {Bland -Hawthorn}, {Brough}, {Colless},
  {Couch}, {Davies}, {Drinkwater}, {Goodwin}, {Hopkins}, {Konstantopoulos},
  {Foster}, {Lawrence}, {Lorente}, {Medling}, {Metcalfe}, {Richards}, {van de
  Sande}, {Scott}, {Shanks}, {Sharp}, {Thomas}, \&
  {Tonini}}]{2017MNRAS.468.1824O}
{Owers}, M.~S., {Allen}, J.~T., {Baldry}, I., {et~al.} 2017,
  \href{http://dx.doi.org/10.1093/mnras/stx562}{\color{magenta}\mnras},
  \href{https://ui.adsabs.harvard.edu/abs/2017MNRAS.468.1824O}{468, 1824}

\bibitem[{{Owers} {et~al.}(2019){Owers}, {Hudson}, {Oman}, {Bland -Hawthorn},
  {Brough}, {Bryant}, {Cortese}, {Couch}, {Croom}, {van de Sande}, {Federrath},
  {Groves}, {Hopkins}, {Lawrence}, {Lorente}, {McDermid}, {Medling},
  {Richards}, {Scott}, {Taranu}, {Welker}, \& {Yi}}]{2019ApJ...873...52O}
{Owers}, M.~S., {Hudson}, M.~J., {Oman}, K.~A., {et~al.} 2019,
  \href{http://dx.doi.org/10.3847/1538-4357/ab0201}{\color{magenta}\apj},
  \href{https://ui.adsabs.harvard.edu/abs/2019ApJ...873...52O}{873, 52}

\bibitem[{{Penoyre} {et~al.}(2017){Penoyre}, {Moster}, {Sijacki}, \&
  {Genel}}]{2017MNRAS.468.3883P}
{Penoyre}, Z., {Moster}, B.~P., {Sijacki}, D., \& {Genel}, S. 2017,
  \href{http://dx.doi.org/10.1093/mnras/stx762}{\color{magenta}\mnras},
  \href{http://adsabs.harvard.edu/abs/2017MNRAS.468.3883P}{468, 3883}

\bibitem[{{Pizzella} {et~al.}(2004){Pizzella}, {Corsini}, {Vega Beltr{\'a}n},
  \& {Bertola}}]{2004A&A...424..447P}
{Pizzella}, A., {Corsini}, E.~M., {Vega Beltr{\'a}n}, J.~C., \& {Bertola}, F.
  2004,
  \href{http://dx.doi.org/10.1051/0004-6361:20047183}{\color{magenta}\aap},
  \href{https://ui.adsabs.harvard.edu/abs/2004A&A...424..447P}{424, 447}

\bibitem[{{Poggianti} {et~al.}(2016){Poggianti}, {Fasano}, {Omizzolo},
  {Gullieuszik}, {Bettoni}, {Moretti}, {Paccagnella}, {Jaff{\'e}}, {Vulcani},
  \& {Fritz}}]{2016AJ....151...78P}
{Poggianti}, B.~M., {Fasano}, G., {Omizzolo}, A., {et~al.} 2016,
  \href{http://dx.doi.org/10.3847/0004-6256/151/3/78}{\color{magenta}\aj},
  \href{https://ui.adsabs.harvard.edu/abs/2016AJ....151...78P}{151, 78}

\bibitem[{{Puerari} \& {Pfenniger}(2001)}]{2001Ap&SS.276..909P}
{Puerari}, I. \& {Pfenniger}, D. 2001,
  \href{http://dx.doi.org/10.1023/A:1017581325673}{\color{magenta}\apss},
  \href{https://ui.adsabs.harvard.edu/abs/2001Ap&SS.276..909P}{276, 909}

\bibitem[{{Quilis} {et~al.}(2000){Quilis}, {Moore}, \&
  {Bower}}]{2000Sci...288.1617Q}
{Quilis}, V., {Moore}, B., \& {Bower}, R. 2000,
  \href{http://dx.doi.org/10.1126/science.288.5471.1617}{\color{magenta}Science},
  \href{http://adsabs.harvard.edu/abs/2000Sci...288.1617Q}{288, 1617}

\bibitem[{{Rhee} {et~al.}(2017){Rhee}, {Smith}, {Choi}, {Yi}, {Jaff{\'e}},
  {Candlish}, \& {S{\'a}nchez-J{\'a}nssen}}]{2017ApJ...843..128R}
{Rhee}, J., {Smith}, R., {Choi}, H., {et~al.} 2017,
  \href{http://dx.doi.org/10.3847/1538-4357/aa6d6c}{\color{magenta}\apj},
  \href{https://ui.adsabs.harvard.edu/abs/2017ApJ...843..128R}{843, 128}

\bibitem[{{Rodriguez-Gomez} {et~al.}(2017){Rodriguez-Gomez}, {Sales}, {Genel},
  {Pillepich}, {Zjupa}, {Nelson}, {Griffen}, {Torrey}, {Snyder},
  {Vogelsberger}, {Springel}, {Ma}, \& {Hernquist}}]{2017MNRAS.467.3083R}
{Rodriguez-Gomez}, V., {Sales}, L.~V., {Genel}, S., {et~al.} 2017,
  \href{http://dx.doi.org/10.1093/mnras/stx305}{\color{magenta}\mnras},
  \href{http://adsabs.harvard.edu/abs/2017MNRAS.467.3083R}{467, 3083}

\bibitem[{{Rubin} {et~al.}(1992){Rubin}, {Graham}, \&
  {Kenney}}]{1992ApJ...394L...9R}
{Rubin}, V.~C., {Graham}, J.~A., \& {Kenney}, J. D.~P. 1992,
  \href{http://dx.doi.org/10.1086/186460}{\color{magenta}\apj},
  \href{https://ui.adsabs.harvard.edu/abs/1992ApJ...394L...9R}{394, L9}

\bibitem[{{Sarzi} {et~al.}(2006){Sarzi}, {Falc{\'o}n-Barroso}, {Davies},
  {Bacon}, {Bureau}, {Cappellari}, {de Zeeuw}, {Emsellem}, {Fathi},
  {Krajnovi{\'c}}, {Kuntschner}, {McDermid}, \&
  {Peletier}}]{2006MNRAS.366.1151S}
{Sarzi}, M., {Falc{\'o}n-Barroso}, J., {Davies}, R.~L., {et~al.} 2006,
  \href{http://dx.doi.org/10.1111/j.1365-2966.2005.09839.x}{\color{magenta}\mnras},
  \href{http://adsabs.harvard.edu/abs/2006MNRAS.366.1151S}{366, 1151}

\bibitem[{{Schaye} {et~al.}(2015){Schaye}, {Crain}, {Bower}, {Furlong},
  {Schaller}, {Theuns}, {Dalla Vecchia}, {Frenk}, {McCarthy}, {Helly},
  {Jenkins}, {Rosas-Guevara}, {White}, {Baes}, {Booth}, {Camps}, {Navarro},
  {Qu}, {Rahmati}, {Sawala}, {Thomas}, \& {Trayford}}]{Eagle}
{Schaye}, J., {Crain}, R.~A., {Bower}, R.~G., {et~al.} 2015,
  \href{http://dx.doi.org/10.1093/mnras/stu2058}{\color{magenta}\mnras},
  \href{http://adsabs.harvard.edu/abs/2015MNRAS.446..521S}{446, 521}

\bibitem[{{Schweizer} {et~al.}(1983){Schweizer}, {Whitmore}, \&
  {Rubin}}]{1983AJ.....88..909S}
{Schweizer}, F., {Whitmore}, B.~C., \& {Rubin}, V.~C. 1983,
  \href{http://dx.doi.org/10.1086/113377}{\color{magenta}\aj},
  \href{https://ui.adsabs.harvard.edu/abs/1983AJ.....88..909S}{88, 909}

\bibitem[{{Serra} {et~al.}(2014){Serra}, {Oser}, {Krajnovi{\'c}}, {Naab},
  {Oosterloo}, {Morganti}, {Cappellari}, {Emsellem}, {Young}, \&
  {Blitz}}]{Serra+14}
{Serra}, P., {Oser}, L., {Krajnovi{\'c}}, D., {et~al.} 2014,
  \href{http://dx.doi.org/10.1093/mnras/stt2496}{\color{magenta}\mnras},
  \href{https://ui.adsabs.harvard.edu/abs/2014MNRAS.444.3388S}{444, 3388}

\bibitem[{{Snyder} {et~al.}(2015){Snyder}, {Torrey}, {Lotz}, {Genel},
  {McBride}, {Vogelsberger}, {Pillepich}, {Nelson}, {Sales}, {Sijacki},
  {Hernquist}, \& {Springel}}]{2015MNRAS.454.1886S}
{Snyder}, G.~F., {Torrey}, P., {Lotz}, J.~M., {et~al.} 2015,
  \href{http://dx.doi.org/10.1093/mnras/stv2078}{\color{magenta}\mnras},
  \href{http://adsabs.harvard.edu/abs/2015MNRAS.454.1886S}{454, 1886}

\bibitem[{{Starkenburg} {et~al.}(2019){Starkenburg}, {Sales}, {Genel},
  {Manzano-King}, {Canalizo}, \& {Hernquist}}]{2019ApJ...878..143S}
{Starkenburg}, T.~K., {Sales}, L.~V., {Genel}, S., {et~al.} 2019,
  \href{http://dx.doi.org/10.3847/1538-4357/ab2128}{\color{magenta}\apj},
  \href{https://ui.adsabs.harvard.edu/abs/2019ApJ...878..143S}{878, 143}

\bibitem[{{Sweet} {et~al.}(2016){Sweet}, {Drinkwater}, {Meurer}, {Kilborn},
  {Audcent-Ross}, {Baumgardt}, \& {Bekki}}]{2016MNRAS.455.2508S}
{Sweet}, S.~M., {Drinkwater}, M.~J., {Meurer}, G., {et~al.} 2016,
  \href{http://dx.doi.org/10.1093/mnras/stv2480}{\color{magenta}\mnras},
  \href{https://ui.adsabs.harvard.edu/abs/2016MNRAS.455.2508S}{455, 2508}

\bibitem[{{Teyssier}(2002)}]{2002A&A...385..337T}
{Teyssier}, R. 2002,
  \href{http://dx.doi.org/10.1051/0004-6361:20011817}{\color{magenta}\aap},
  \href{http://adsabs.harvard.edu/abs/2002A%26A...385..337T}{385, 337}

\bibitem[{{Thakar} \& {Ryden}(1996)}]{1996ApJ...461...55T}
{Thakar}, A.~R. \& {Ryden}, B.~S. 1996,
  \href{http://dx.doi.org/10.1086/177037}{\color{magenta}\apj},
  \href{https://ui.adsabs.harvard.edu/abs/1996ApJ...461...55T}{461, 55}

\bibitem[{{Torrey} {et~al.}(2012){Torrey}, {Vogelsberger}, {Sijacki},
  {Springel}, \& {Hernquist}}]{2012MNRAS.427.2224T}
{Torrey}, P., {Vogelsberger}, M., {Sijacki}, D., {Springel}, V., \&
  {Hernquist}, L. 2012,
  \href{http://dx.doi.org/10.1111/j.1365-2966.2012.22082.x}{\color{magenta}\mnras},
  \href{http://adsabs.harvard.edu/abs/2012MNRAS.427.2224T}{427, 2224}

\bibitem[{{Tweed} {et~al.}(2009){Tweed}, {Devriendt}, {Blaizot}, {Colombi}, \&
  {Slyz}}]{2009A&A...506..647T}
{Tweed}, D., {Devriendt}, J., {Blaizot}, J., {Colombi}, S., \& {Slyz}, A. 2009,
  \href{http://dx.doi.org/10.1051/0004-6361/200911787}{\color{magenta}\aap},
  \href{http://adsabs.harvard.edu/abs/2009A%26A...506..647T}{506, 647}

\bibitem[{{Ulrich}(1975)}]{1975PASP...87..965U}
{Ulrich}, M.-H. 1975,
  \href{http://dx.doi.org/10.1086/129881}{\color{magenta}\pasp},
  \href{https://ui.adsabs.harvard.edu/abs/1975PASP...87..965U}{87, 965}

\bibitem[{{van de Sande} {et~al.}(2019){van de Sande}, {Lagos}, {Welker},
  {Bland-Hawthorn}, {Schulze}, {Remus}, {Bah{\'e}}, {Brough}, {Bryant},
  {Cortese}, {Croom}, {Devriendt}, {Dubois}, {Goodwin}, {Konstantopoulos},
  {Lawrence}, {Medling}, {Pichon}, {Richards}, {Sanchez}, {Scott}, \&
  {Sweet}}]{2019MNRAS.484..869V}
{van de Sande}, J., {Lagos}, C. D.~P., {Welker}, C., {et~al.} 2019,
  \href{http://dx.doi.org/10.1093/mnras/sty3506}{\color{magenta}\mnras},
  \href{https://ui.adsabs.harvard.edu/abs/2019MNRAS.484..869V}{484, 869}

\bibitem[{{van de Voort} {et~al.}(2015){van de Voort}, {Davis}, {Kere{\v{s}}},
  {Quataert}, {Faucher-Gigu{\`e}re}, \& {Hopkins}}]{2015MNRAS.451.3269V}
{van de Voort}, F., {Davis}, T.~A., {Kere{\v{s}}}, D., {et~al.} 2015,
  \href{http://dx.doi.org/10.1093/mnras/stv1217}{\color{magenta}\mnras},
  \href{https://ui.adsabs.harvard.edu/abs/2015MNRAS.451.3269V}{451, 3269}

\bibitem[{{van Gorkom} {et~al.}(1987){van Gorkom}, {Schechter}, \&
  {Kristian}}]{1987ApJ...314..457V}
{van Gorkom}, J.~H., {Schechter}, P.~L., \& {Kristian}, J. 1987,
  \href{http://dx.doi.org/10.1086/165078}{\color{magenta}\apj},
  \href{https://ui.adsabs.harvard.edu/abs/1987ApJ...314..457V}{314, 457}

\bibitem[{{Vogelsberger} {et~al.}(2014){Vogelsberger}, {Genel}, {Springel},
  {Torrey}, {Sijacki}, {Xu}, {Snyder}, {Nelson}, \& {Hernquist}}]{Illustris}
{Vogelsberger}, M., {Genel}, S., {Springel}, V., {et~al.} 2014,
  \href{http://dx.doi.org/10.1093/mnras/stu1536}{\color{magenta}\mnras},
  \href{http://adsabs.harvard.edu/abs/2014MNRAS.444.1518V}{444, 1518}

\bibitem[{{Whitmore} {et~al.}(1990){Whitmore}, {Lucas}, {McElroy},
  {Steiman-Cameron}, {Sackett}, \& {Olling}}]{1990AJ....100.1489W}
{Whitmore}, B.~C., {Lucas}, R.~A., {McElroy}, D.~B., {et~al.} 1990,
  \href{http://dx.doi.org/10.1086/115614}{\color{magenta}\aj},
  \href{https://ui.adsabs.harvard.edu/abs/1990AJ....100.1489W}{100, 1489}

\end{thebibliography}

\appendix
\section{The grid-locking problem}
\label{sec:AppA}

The angular momenta of stars and the gas in the grid-based simulation tend to align with the directions of grids; the X, Y, and Z-axes of the simulation box effect \citep{2015MNRAS.454.2736C, 2020MNRAS.tmp..237K}. We discuss the impact of this ``grid-locking'' on our study. In conclusion, we found that grid-locking effect is present in Horizon-AGN, but it does not affect the main results. 

\restartappendixnumbering
\begin{figure}
	\includegraphics[width=\columnwidth]{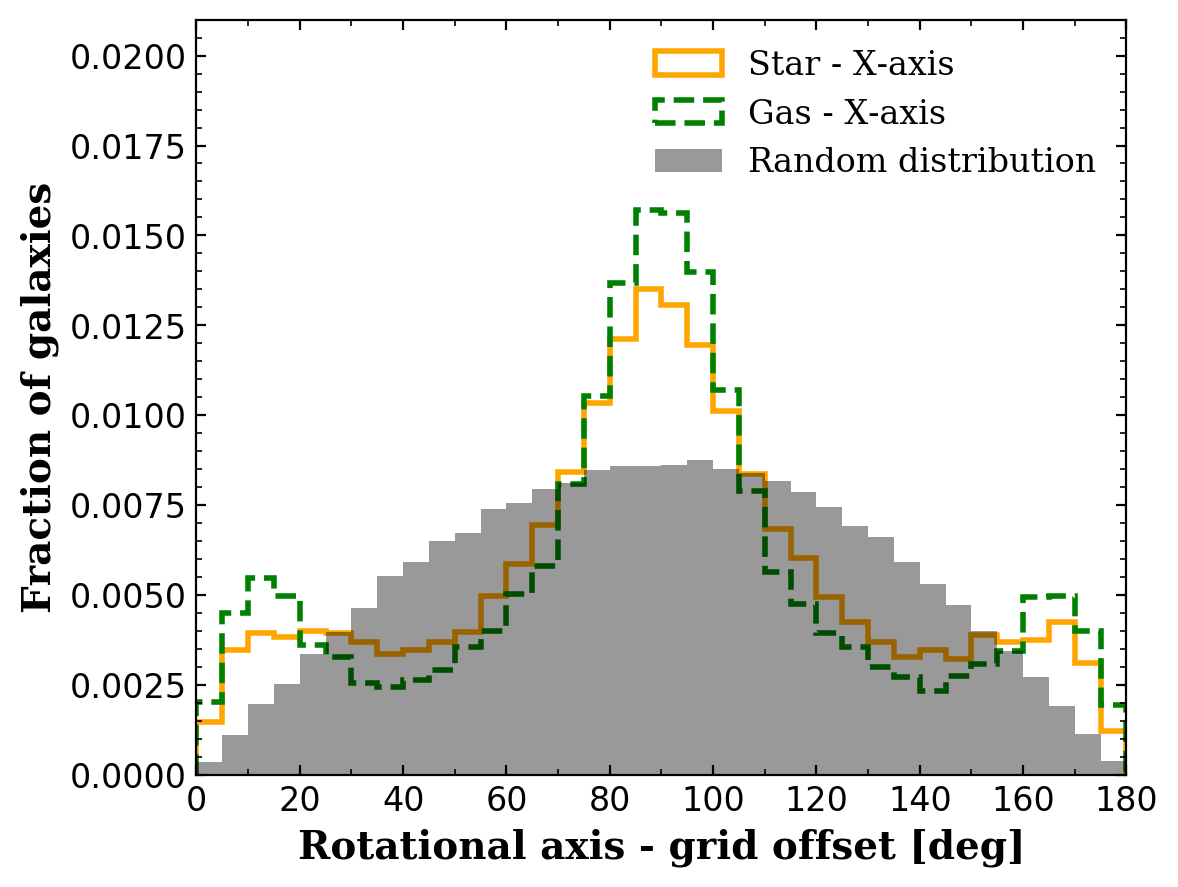}
    \caption{The distribution of the angle offset between a Cartesian grid (X) and rotational axes (star and gas). Compare with the distribution of the randomly distributed galaxies (gray), there are more Horizon-AGN galaxies with angle offsets of stellar (yellow) and gas (green) rotational axes at around 0, 90 or 180 degrees. We note that all three axes (X, Y, and Z) show virtually the same phenomenon.}
    \label{fig:grid}
\end{figure}

\begin{figure*}
	\includegraphics[width=\linewidth]{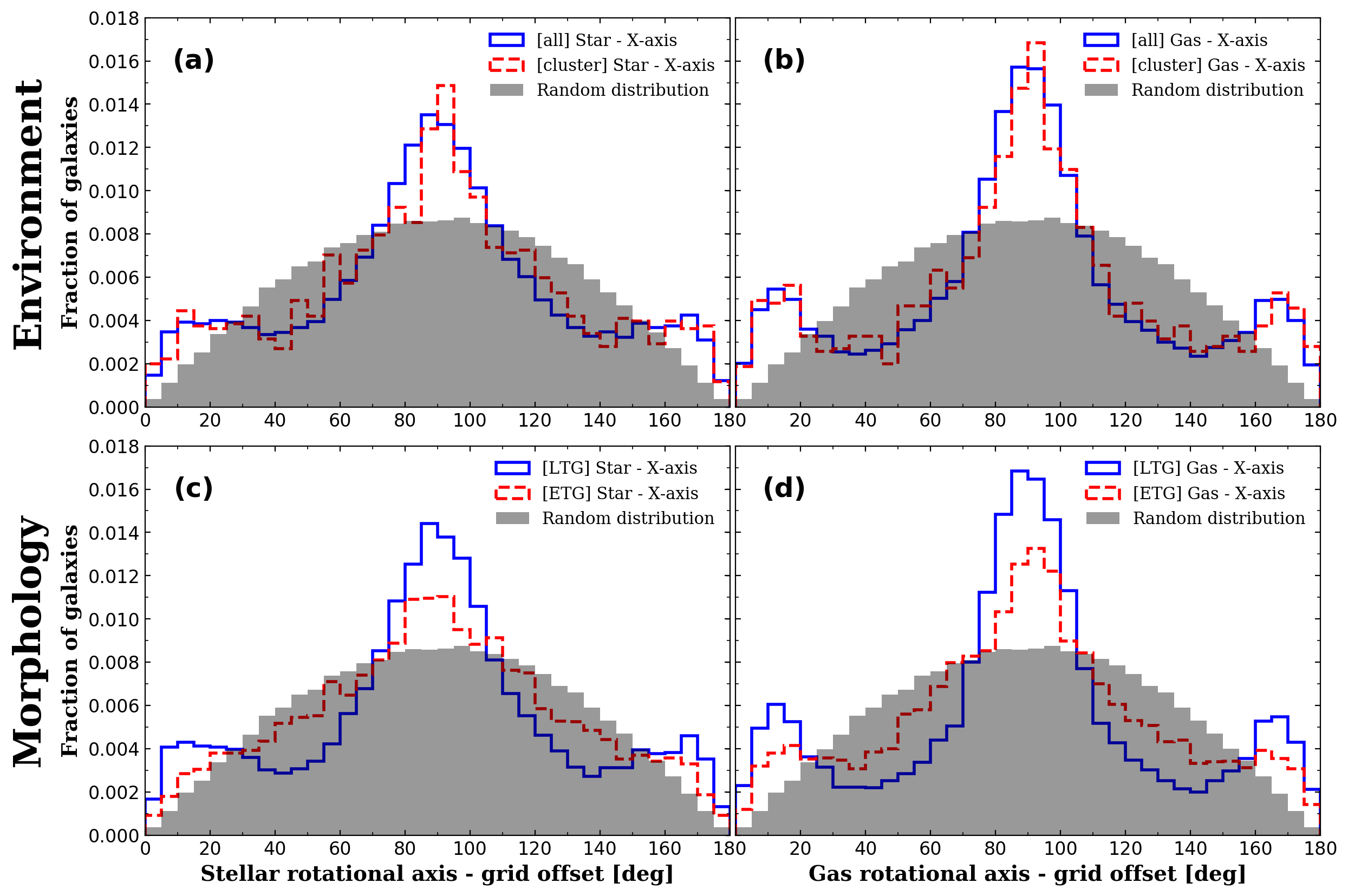}
    \caption{The distribution of the angle offset between a Cartesian grid (X) and the rotational axes (star and gas) for the different samples. For the comparison, the distribution of the randomly distributed galaxies is shown as a gray histogram. We note that all three axes (X, Y, and Z) show virtually the same phenomenon. While the shapes of the histograms of the whole and cluster galaxies are very similar to each other, LTGs seem to be more affected by grid-locking than ETGs. (a): the angle offset between a grid and the stellar rotational axes for the whole and cluster galaxies (b): same as (a) but for gas (c): the angle offset between a grid and the stellar rotational axes for ETGs and LTGs (d): same as (c) but for gas.}
    \label{fig:grid2}
\end{figure*}

\begin{figure*}
	\includegraphics[width=\linewidth]{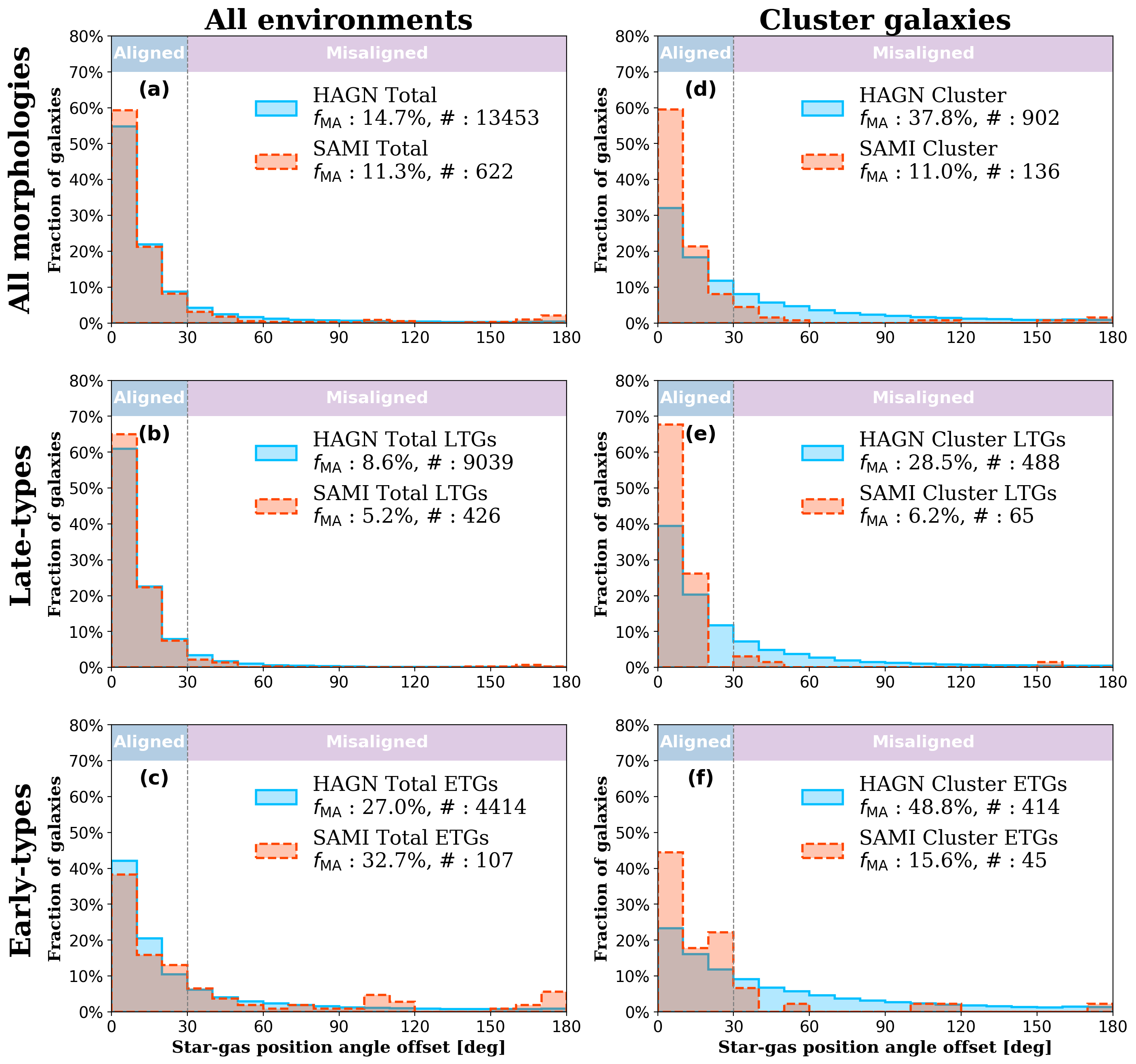}
    \caption{The distribution of star-gas PA offset in the grid-locking safe Horizon-AGN (blue) and SAMI (red) galaxies. We define a ‘grid-locking safe’ subsample of galaxies by choosing only the galaxies whose gas angular momentum axis is at least 20 degrees away from all three grid axes. We have performed a Monte-Carlo simulation of projection on the Horizon-AGN galaxies as many as 1,000 times to minimize bias from the projection effect (see Fig.~\ref{fig:image_SAMI_dist}).}
    \label{fig:grid3}
\end{figure*}

Fig.~\ref{fig:grid} shows the distribution of the angle offset between grids and rotational axes (star and gas). When galaxies are randomly distributed inside a Cartesian coordinate, their axes of angular momenta have the highest probability of being offset by 90 degrees from any of the three Cartesian axes, as illustrated by the gray histogram. We find that both the stellar and gas angular momenta of Horizon-AGN galaxies preferentially align with the direction of grids. Compared with the random distribution, there are more galaxies with angle offsets of stellar/gas rotational axes at around 0, 90 or 180 degrees. This is an artifact of the AMR technique and known as grid-locking. The grid-locking effect is more pronounced for gas than for stars. Fig.~\ref{fig:grid} shows the offset distribution of stars and gas with respect to one (X) axis, while all three axes (X, Y, and Z) show virtually the same phenomenon. 

It is important to note that our main concern is the relative misalignment fractions between the galaxy populations of different morphologies, gas contents, and environments. If the impact of grid-locking affects all galaxies by the same degree, any bias owing to the grid-locking would cancel out. Figs.~\ref{fig:grid2}-(a) and -(b) compare the distributions of the angle offset between the grids and the rotational axes of the whole (blue) and cluster (red) galaxies. The shapes of the histograms are very similar to each other, suggesting that the difference in misalignment fractions between the galaxies in the whole and cluster environments (see Section~\ref{sec:dis_env}) is not originating from the grid-locking effect. 
 
However, we have found that LTGs are more affected by grid-locking than ETGs, as shown in Figs.~\ref{fig:grid2}-(c) and -(d). LTGs tend to have higher gas fractions so that they are more sensitive to the direction of the grid. 
Note that this morphology dependence of the grid-locking effect affects the overall misalignment fraction.
To investigate the effect further, we define a ‘grid-locking safe’ subsample of galaxies by choosing only the galaxies whose gas angular momentum axis is at least 20 degrees away from all three grid axes. Fig.~\ref{fig:grid3} shows the same as Fig.~\ref{fig:image_SAMI_dist} (of the main text) but for the grid-locking safe galaxies. The use of the grid-locking safe sample increased the misalignment fractions in all panels as expected, but our main conclusion remains the same; LTGs tend to be more aligned than ETGs.

Grid-locking works in a way like friction to misalignment, but it is difficult at the moment to understand whether and how it cumulates over time. A future investigation using different grid resolutions may provide a hint to this question.

\bibliographystyle{aasjournal}



\end{document}